\documentclass[a4paper,11pt]{article}
\usepackage{jcappub}

\usepackage{mathtools}
\usepackage{orcidlink}
\usepackage[T1]{fontenc}
\usepackage{lmodern}
\usepackage[makeroom]{cancel}
\usepackage[titletoc]{appendix}
\usepackage{hyperref}
\usepackage{csquotes}
\usepackage{slashed}
\usepackage{adjustbox}
\usepackage{xcolor}
\usepackage{dcolumn}
\usepackage{amsfonts}
\usepackage{indentfirst}
\usepackage{afterpage}
\usepackage{graphicx}
\usepackage{rotating}
\usepackage{amssymb}
\usepackage{braket}
\usepackage{mathrsfs}
\usepackage{subcaption}
\usepackage{appendix}
\usepackage{natbib}
\usepackage{float}
\usepackage{color}
\usepackage{soul}
\usepackage{microtype}
\usepackage{enumerate}
\usepackage{physics}
\usepackage{comment}
\usepackage{academicons}
\usepackage{multirow}

\newcommand{\beq}{\begin{equation}}
\newcommand{\eeq}{\end{equation}}
\newcommand{\ba}{\begin{eqnarray}}
\newcommand{\ea}{\end{eqnarray}}
\newcommand{\alp}{\alpha}

\newcommand{\eps}{\epsilon}
\newcommand{\kap}{\kappa}

\newcommand{\bfs}[1]{\boldsymbol{#1}}

\usepackage{graphicx}
\graphicspath{{./Figures/}}
\usepackage[margin=15pt,font={sf,it},labelfont={bf,sf},format=plain,justification=centerlast,width=.9\textwidth,skip=8pt]{caption}
\usetikzlibrary{matrix,calc,decorations.markings}
\usepackage{amsmath,amsthm}\usepackage[scaled=1.,ncf,vvarbb]{newtxmath}
\definecolor{DarkViolet}{RGB}{148,0,211}
\definecolor{DarkBlue}{RGB}{0,0,154}

\newcommand{\mycolor}{DarkViolet}
\definecolor{tclr}{RGB}{148,0,211}

\hypersetup{colorlinks=true,bookmarksopen,bookmarksnumbered,citecolor={\mycolor},linkcolor={DarkBlue},urlcolor={DarkBlue},pdfstartview=FitH,anchorcolor={DarkBlue}}

\setstcolor{red}

\title{\huge{Dirty Black Holes, Clean Signals: Near-Horizon vs.~Environmental Effects on Grey-Body Factors and Hawking Radiation}}

\author[a]{Roman A.~Konoplya~\orcidlink{0000-0003-1343-9584}}
\emailAdd{roman.konoplya@gmail.com}

\author[a]{and Thomas D.~Pappas~\orcidlink{0000-0003-2186-357X}}
\emailAdd{thomas.pappas@physics.slu.cz}

\affiliation[a]{\href{https://www.slu.cz/phys/en/}{Research Centre for Theoretical Physics and Astrophysics,\\ Institute of Physics, Silesian University in Opava},\\ Bezručovo náměstí 13, CZ-746 01 Opava, Czech Republic}

\abstract{Grey-body factors are not only essential ingredients for computing the intensity of Hawking radiation, but also serve as  characteristics of black hole's geometry that are closely related to their quasinormal modes. Importantly, they tend to be more stable under small deformations of the background spacetime. In this work, we carry out a detailed analysis of grey-body factors and Hawking radiation for a spherically symmetric black hole subject to localized deformations which do not alter the Hawking temperature: near-horizon modifications to simulate possible new physics or matter fields, and far-zone perturbations to model environmental or astrophysical effects. We show that environmental deformations have only a minor impact on the grey-body factors and Hawking radiation--unless the additional potential barrier created by the environment becomes comparable in height to the primary peak associated with the black hole itself, a scenario more relevant to nonlinear dynamics. In contrast, near-horizon deformations significantly affect the Hawking spectrum, particularly in the low-frequency regime.}

\begin{document}
\maketitle
\hypersetup{urlcolor={\mycolor}}
\flushbottom

\section{Introduction}
\label{Sec:Introduction}

Grey-body factors (GFs) play a central role in black hole physics, as they determine the frequency-dependent probability for Hawking radiation to escape to infinity after being partially scattered by the black hole's potential barrier.~However, their significance extends beyond black hole thermodynamics \cite{Oshita:2023cjz}. Recent developments have revealed a correspondence between grey-body factors \cite{Page:1974he,Page:1976ki}  and quasinormal modes (QNMs) \cite{Kokkotas:1999bd,Konoplya:2011qq,Bolokhov:2025uxz} which was first conjectured in \cite{Kyutoku:2022gbr,Oshita:2023cjz} and later formulated more precisely in \cite{Konoplya:2024lir,Konoplya:2024vuj},  the latter being key signatures in gravitational wave observations. While QNMs are sensitive probes of the near-horizon structure and can exhibit dramatic changes under small geometric deformations, grey-body factors typically respond more smoothly to such modifications \cite{Rosato:2024arw,Oshita:2024fzf,Wu:2024ldo}. This makes them a particularly stable spectral characteristic of black hole spacetimes — valuable both for modeling Hawking radiation \cite{Page:1976df,Page:1976ki,Kanti:2004nr} and for cross-validating gravitational wave spectrum in the context of modified gravity or exotic compact objects~\cite{Rosato:2025byu,Pedrotti:2025upg,Malik:2024cgb,Tang:2025mkk,Lutfuoglu:2025hjy,Konoplya:2025hgp,Bolokhov:2024otn,Lutfuoglu:2025ohb}.

The latter is possible because, although quasinormal modes and grey-body factors correspond to different boundary conditions, there exists the aforementioned correspondence between them. In particular, this correspondence holds for the part of the quasinormal spectrum that can be well approximated by the WKB method developed for a single maximum effective potential in \cite{Schutz:1985km,Iyer:1986np,Konoplya:2003ii,Matyjasek:2017psv}, and generally breaks down for more complicated effective potentials. Even when the correspondence is valid, it is exact only in the eikonal regime and becomes approximate and less accurate beyond it \cite{Malik:2024cgb,Bolokhov:2024otn,Lutfuoglu:2025ljm,Lutfuoglu:2025hjy,Dubinsky:2024nzo,Bolokhov:2025lnt,Dubinsky:2025ypj,Dubinsky:2025nxv,Lutfuoglu:2025eik,Lutfuoglu:2025blw}. Naturally, higher overtones, which are poorly captured by the WKB method, are largely irrelevant for the correspondence. Therefore, the behavior of grey-body factors in response to the detailed shape of the effective potential is not, strictly speaking, directly linked to that of quasinormal modes, especially when additional features such as bumps or dips arise due to new physics near the horizon or due to astrophysical environments at large distances.

While grey-body factors have been computed for a wide variety of black hole models — including those with higher curvature corrections, exotic matter content, or localized perturbations \cite{Yuan:2025eyi,Dubinsky:2024vbn,Konoplya:2019hml,Konoplya:2019ppy,Kanti:2008eq,Grain:2005my,Konoplya:2020cbv,Konoplya:2020jgt,Dubinsky:2024nzo,Konoplya:2010vz,Konoplya:2010kv,Malik:2024wvs,Dubinsky:2025fwv,Bonanno:2025dry} — there remains a lack of a systematic study assessing the distinct impact of near-horizon and far-zone deformations on these quantities. In particular, it is not yet well understood how such geometric modifications influence not only the grey-body factors themselves, but also physically measurable, gauge-invariant observables such as the total emission rates of Hawking radiation. Addressing this gap, the present work investigates how both types of deformations—those localized near the event horizon and those situated far from the black hole—affect the transmission of radiation and the overall spectral intensity. By doing so, we aim to clarify the behavior of grey-body factors and Hawking fluxes in the presence of "dirty" black holes, that is, spacetimes embedded in complex  environments.

Previous studies of GFs and QNMs with theory-agnostic modifications of GR are mainly dedicated either to highly localized deformations of the geometry in the whole space outside the event horizon \cite{Oshita:2024fzf} or various parametrized black hole metrics \cite{Dubinsky:2024nzo,Dubinsky:2024rvf,Thomopoulos:2025nuf}. In either case no Hawking radiation has been considered. Quasinormal modes of black holes with physically motivated deformations of the geometry localized either near the event horizon or in the far zone were considered in~\cite{Konoplya:2022pbc}. It was shown that, while the fundamental mode is largely insensitive to near-horizon deformations, the overtones exhibit a strong dependence on them, with deviations from the Schwarzschild values growing rapidly with increasing overtone number. Furthermore, in~\cite{Konoplya:2024wds}, it was demonstrated that such near-horizon bumps also affect the slowly decaying oscillatory late-time tails of massive fields.

Thus, while there exists a relatively extensive body of literature and at least a general understanding of how quasinormal spectra are affected by deformations of the background geometry — whether due to new physics near the event horizon or modifications in the black hole environment at large distances — the same level of understanding has not yet been achieved for grey-body factors and Hawking radiation. In this work, we aim to address this gap by studying grey-body factors and Hawking radiation of the Schwarzschild black hole in the presence of various types of metric deformations (``bumps'' and ``dips'') and the corresponding modifications to the effective potential.

We will show that while both near-horizon and far-zone deformations of the effective potential influence grey-body factors and Hawking radiation, the magnitude of the effect is strongly dependent on the location and amplitude of the deformation. Near-horizon modifications, even when small, can significantly enhance or suppress the total emissivity —especially for low multipole numbers — because they directly alter the main barrier structure governing wave transmission. In contrast, far-zone environmental perturbations have a negligible impact unless they are unphysically large, confirming that typical astrophysical matter distributions such as accretion disks are unlikely to measurably modify Hawking fluxes. Our analysis includes a detailed numerical study of Maxwell and Dirac fields, with a comparison of different bump profiles, and explores general patterns of emission rate variation under changes to bump height, width, and position.

The paper is organized as follows. In Section~\ref{Sec:GFs_and_HR}, we briefly review the formalism of grey-body factors and Hawking radiation in spherically symmetric spacetimes, focusing on the propagation of Maxwell and Dirac fields. Section~\ref{Sec:Near_horizon_deformations} is devoted to analyzing near-horizon deformations of the effective potential and their impact on grey-body spectra and emission rates. In Section~\ref{Sec:Far_zone_deformations}, we consider deformations arising from the far-zone environment and study their relative influence. In Section~\ref{Sec:General_laws}, we derive general trends for the total emissivity of the black hole under various deformation scenarios. In Section~\ref{Sec:Generalization}, we outline possible directions in which our analysis can be generalized. Finally, Section~\ref{Sec:Conclusions} summarizes our main findings and discusses possible extensions of this work.

\section{Grey-body factors and Hawking radiation}
\label{Sec:GFs_and_HR}

The line element for a general static and spherically symmetric black hole space-time described in terms of a single metric function $f(r)$, can be written as
\beq
\dd s^2=-f(r) \dd t^2+\frac{\dd r^2}{f(r)}+r^2 \left( \dd \theta^2+\sin^2\theta \dd \phi^2 \right)\,,
\label{eq:ds2}
\eeq
with the radius $r_0$ of the outer event horizon determined by $f(r_0)=0$.

In our analysis, we will be constrained to particles that can be emitted during the semi-classically described evaporation process~\cite{Hawking:1975vcx}. In this context, considering the emission of Maxwell and Dirac fields is sufficient to capture the essential features of the evaporation process. The emission of gravitons is typically strongly suppressed compared to that of matter fields~\cite{Page:1976df,Page:1977um}, at least in the case of four-dimensional black holes.

The general covariant equations governing the propagation of the electromagnetic field $A_\mu$ and the massless Dirac field $Y$ on the curved background~\eqref{eq:ds2} are respectively the following
\ba
\frac{1}{\sqrt{-g}}\partial_{\mu}\left(F_{\rho\sigma}g^{\rho\nu}g^{\sigma\mu}\sqrt{-g} \right)&=&0\,,\label{eq:em_field_eq}\\
\gamma^{\alp}\left(\frac{\partial}{\partial^{\alp}}-\Gamma_{\alp}
\right)Y&=&0\label{eq:Dirac_field_eq}\,,
\ea
where the electromagnetic tensor is given by $F_{\mu\nu}=\partial_{\mu}A_{\nu}-\partial_{\nu}A_{\mu}$, gamma matrices in curved space-time are denoted by $\gamma^{\alp}$, and $\Gamma_{\alp}$ are spin connections in the tetrad formalism. Upon performing separation of variables and transitioning to the so-called tortoise coordinate defined via
\beq
\frac{\dd r_*}{\dd r}=\frac{1}{f(r)}\,,
\eeq
the field equations~\eqref{eq:em_field_eq} and~\eqref{eq:Dirac_field_eq}, can be recast to the Sch\"odinger-like form
\beq
\frac{\dd^2 \Psi}{\dd r_*^2} +\left[\omega^2-V(r) \right] \Psi=0\,.
\label{eq:wave_eq}
\eeq
In the case of the electromagnetic field, the effective potential is given by
\beq
V_{em}(r)=f(r)\frac{l (l+1)}{r^2}\,,
\label{eq:Veff_em}
\eeq
where $l=1,2,3,\ldots$ is the multipole number. On the other hand, for the Dirac field there exist two isospectral potentials given by
\beq
V_{\pm}(r)=\frac{f(r) }{r^2}\left(l+\frac{1}{2} \right)^2\pm \left(l+\frac{1}{2} \right) \frac{\dd}{\dd r_*}\frac{\sqrt{f(r)}}{r}\,,
\label{eq:Veff_Dirac}
\eeq
where now, the fermionic multipole number takes the values $l=1/2,3/2,5/2,\ldots$ Due to isospectrality we consider only $V_{+}(r)$ for our analysis.

The accurate determination of the grey-body factors requires the numerical integration of the wave equation~\eqref{eq:wave_eq}, subject to boundary conditions corresponding to purely in-going modes at the black-hole horizon and both in-going and out-going modes at spatial infinity. The tortoise coordinate, maps the coordinate location of $r_0$ to $r_* \to -\infty$ and $r \to \infty$ to $r_* \to \infty$ respectively, where the effective potentials~\eqref{eq:Veff_em} and~\eqref{eq:Veff_Dirac} satisfy
\beq
\lim_{r_*\to\pm\infty}V_{eff}\left(r(r_*)\right)=0\,.
\eeq
As such, the general approximate analytic solution to~\eqref{eq:wave_eq} near the BH horizon can be written as
\beq
\Psi_0=B^{in}_0 e^{-i\omega r_*}+B^{out}_0 e^{i\omega r_*}\,,
\label{eq:Psi_0}
\eeq
and at spatial infinity as
\beq
\Psi_\infty=B^{in}_\infty e^{-i\omega r_*}+B^{out}_\infty e^{i\omega r_*}\,,
\label{eq:Psi_c}
\eeq
where $B^{in}_{(0,\infty)}$ and $B^{out}_{(0,\infty)}$ are the amplitudes for in-going and out-going waves respectively when $\omega>0$. The requirement of purely in-going waves at the BH horizon, amounts to setting $B^{out}_0=0$ in~\eqref{eq:Psi_0} and subsequently, for any given value of $\omega$, by numerically integrating~\eqref{eq:wave_eq}, one is able to determine the corresponding amplitudes $B^{in}_\infty$ and $B^{out}_\infty$ by comparison of~\eqref{eq:Psi_c} with the numerical solution. Then, at any given frequency $\omega$, the  GFs can be evaluated in terms of $B^{out}_\infty$ and $B^{in}_\infty$ as follows
\beq
|A|^2\equiv1-\left| \frac{B^{out}_\infty}{B^{in}_\infty} \right|^2\,.
\eeq

For the purposes of our study, the GFs have to be calculated with high precision, and we made sure that our numerical GF data are accurate to 14 decimals. In order to achieve this we followed the method described in~\cite{Konoplya:2024vuj}. In particular, for the integration of~\eqref{eq:wave_eq} we imposed initial conditions at the radius $r_i=r_0+\eps$ with
\beq
0<\frac{\eps}{r_0}\ll1\,,
\eeq
given by a series expansion of the ingoing wave at the event horizon
\beq
\Psi(r)=\left(r-r_0\right)^{-i\omega r_0}\sum_{n=0}^{\infty} c_n \left(r-r_0\right)^n\,,
\eeq
and we have included sufficient number of coefficients to achieve the desired level of precision. Furthermore, in order to determine $B^{out}_\infty$  and $B^{in}_\infty$ accurately, we performed a fitting of the numerical solution in the regime $r/r_0\gg1$ in terms of a superposition of expansions of ingoing and outgoing waves as
\beq
\Psi(r)=B^{in}_\infty e^{i\omega r}r^\alpha \sum_{n=0}^{\infty}\frac{F_n}{r^n}+B^{out}_\infty e^{-i\omega r}r^{-\alpha} \sum_{n=0}^{\infty}\frac{G_n}{r^n}\,,
\eeq
and, once again, sufficient number of coefficients had to be taken into account.

Once the accurate values for the GFs have been obtained at each frequency $\omega$, the differential energy emission rate (EER) and total emissivity for Hawking radiation~\cite{Hawking:1975vcx} can then be computed respectively via the formulas
\ba
\frac{\dd^2E}{\dd t\,\dd\omega}=\sum_{l}\frac{N_l |A_l|^2}{2\pi}\frac{\omega}{e^{\omega/T_H}\pm1}\,,\label{eq:EER_def}\\
\frac{\dd E}{\dd t}=\sum_{l}\int{}\frac{N_l |A_l|^2}{2\pi}\frac{\omega}{e^{\omega/T_H}\pm1}\dd \omega\,,
\label{eq:totEER_def}
\ea
where $|A_l|^2$ is the GF for the multipole number $l$, while the multiplicity of states is given by $N_l=2(2l+1)$ for the electromagnetic field, and $N_l=36(l+1/2)$ for the Dirac field~(see e.g.~\cite{Bolokhov:2024voa}). The Hawking temperature $T_H$ in~\eqref{eq:EER_def}-\eqref{eq:totEER_def}, can be computed in terms of the surface gravity on the event horizon of the black hole and for the line-element~\eqref{eq:ds2} it is given by~\cite{York:1984wp}
\beq
T_H=\lim_{r\to r_0}\frac{|f'(r)|}{4 \pi}\,.
\label{eq:T_H}
\eeq
When computing the energy emission rates, we assume that during a sufficiently short time interval — between the emission of two consecutive particles — the black hole temperature remains effectively constant. Due to~\eqref{eq:T_H}, this also corresponds to neglecting any backreaction on the spacetime geometry, allowing thus one to use~\eqref{eq:EER_def}-\eqref{eq:totEER_def} without modifications~\cite{Parikh:1999mf}. In this approximation, the emitted particles are treated as forming a canonical ensemble.

\begin{figure*}[h!]
\begin{center}
\includegraphics[width=0.49\linewidth]{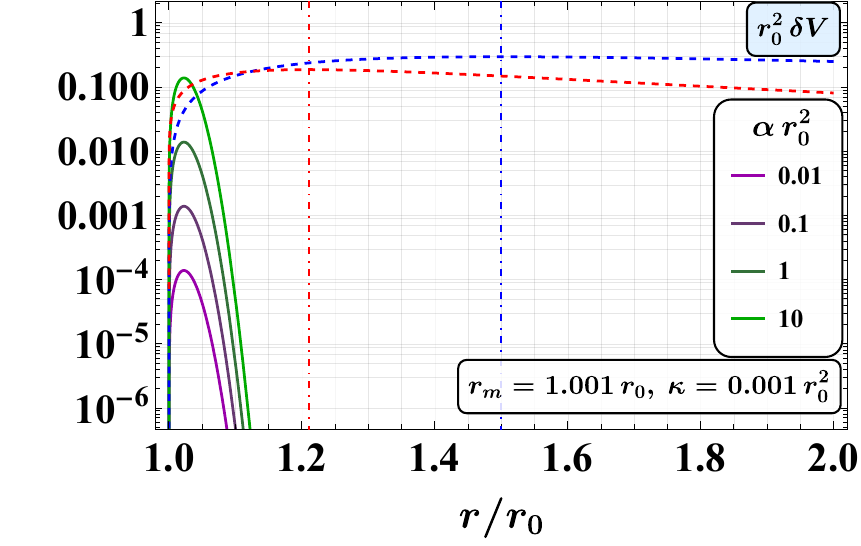}
\includegraphics[width=0.49\linewidth]{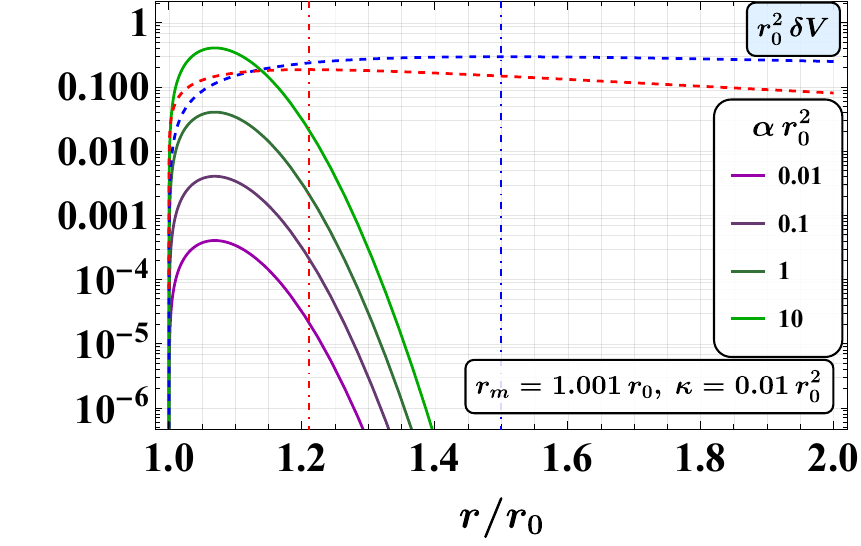}
\includegraphics[width=0.49\linewidth]{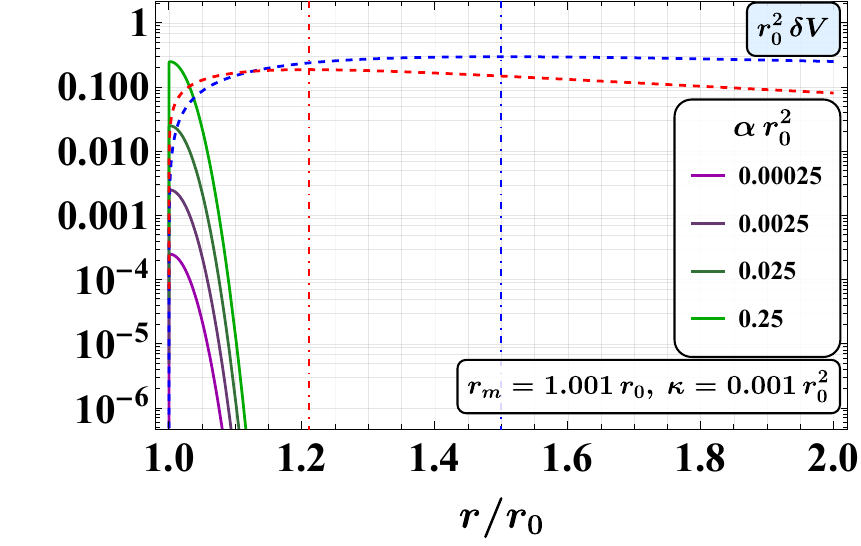}
\includegraphics[width=0.49\linewidth]{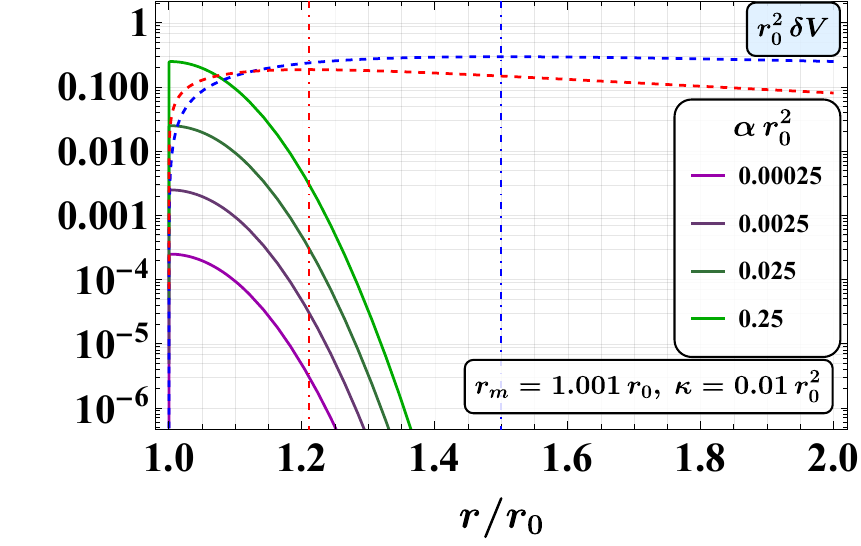}
\caption{Near-horizon Gaussian bump deformations (top panels:~\eqref{eq:Gaussian_bump_typical}, bottom panels:~\eqref{eq:Gaussian_bump_eps}) of the effective potentials on the Schwarzschild background for various values of the height parameter $\alpha$ and two different values of the width parameter (left and right panels). For reference, we also depict the effective potentials for the electromagnetic field with $l=1$ (blue) and the Dirac field with $l=1/2$ (red) with the dashed curves, while the locations of their peaks are indicated by the dot-dashed lines respectively.}
\label{fig:Gaussian_bumps}
\end{center}
\end{figure*}

In order to simulate strong-gravity deviations from GR near the horizon, and effects from the astrophysical environment of the black hole in a model-agnostic way, we introduce localized deformations of the effective potential in~\eqref{eq:wave_eq} according to
\beq
V(r) \to \widetilde{V}(r) = V(r) + \delta V(r)\,,
\label{eq:Gaussian_bump}
\eeq
where $\widetilde{V}(r)$ is the deformed effective potential. The deformation function $\delta V(r)$ in~\eqref{eq:Gaussian_bump}, is typically parametrized by a Gaussian bump of the form~\cite{Konoplya:2024wds}
\beq
\delta V(r) = \alpha \, f(r) e^{-(r-r_m)^2/\kap}\,,
\label{eq:Gaussian_bump_typical}
\eeq
where the parameters $\alpha$, $r_m$, and $\kappa$ are associated with the height, location, and width of the deformation respectively, and $\delta V(r_0)=0$ is ensured by construction. The location of the peak of the deformation~\eqref{eq:Gaussian_bump_typical} corresponds to the solution of
\beq
\frac{\dd \delta V(r)}{\dd r} =0 \quad \Rightarrow \quad f'(r)=\frac{2(r-r_m)}{\kap}f(r)\,,
\label{eq:deltaV_extr}
\eeq
and it is not identified with $r_m$. In fact, the closer $r_m$ is to the horizon of the black hole, the more $f(r)$ deviates from the asymptotically flat limit where $f\to1$ and $f' \to 0$ and as a consequence of~\eqref{eq:deltaV_extr}, the more $r_m$ deviates from the actual location of the peak of the deformation. This discrepancy means that for fixed $\alpha$ and $\kappa$, different values of $r_m$ result to deformations with different heights. Furthermore, for fixed values of $\alpha$ and $r_m$, different values of $\kappa$ modify significantly the location (and height) for the peak of the deformation, when $r_m$ is near the horizon of the black hole, see top panels in Fig.~\ref{fig:Gaussian_bumps}.

From the above, it becomes clear that for the purposes of identifying the effect that each of the three main features of the deformation, namely height, width and location have on the grey-body factors and Hawking radiation, the parametrization~\eqref{eq:deltaV_extr} is inadequate. To this end, we will work instead with the following Gaussian bump
\beq
\delta V(r) = \alpha \, \frac{f(r)}{\eps+f(r)} e^{-(r-r_m)^2/\kap}\,,
\label{eq:Gaussian_bump_eps}
\eeq
which, for sufficiently small values of the dimensionless parameter $\eps$ (that we fix to $\eps=10^{-10}$ throughout the article), provides a deformation function that is free of the aforementioned issues\footnote{In principle, these issues can also be addressed simply by setting $f(r)=1$ in~\eqref{eq:Gaussian_bump_typical} and working with parameter values such that $\delta V(r_0)$ is less than the working numerical precision, thus effectively ensuring that $\delta V(r_0)=0$ for the computations. This approach however requires extremely small values for $\kappa$ when $r_m$ is near $r_0$ in order to avoid significant modifications to the main peak of the effective potentials, and so the range of bump widths that can be considered for the study gets strongly restricted.}, see Fig.~\ref{fig:Gaussian_bumps}. The height of the deformations can be taken to be either positive (bump) or negative (dip), and as far as the modeling of the black-hole environment is concerned, a dip corresponds to normal matter, while a bump represents the presence of phantom matter~\cite{Konoplya:2018yrp}.

Finally, let us close this section with a brief remark. In the original derivation of the Hawking effect~\cite{Hawking:1975vcx}, Hawking based his analysis on wave-packets. In the framework of this approach, one may be concerned about the preservation of the shape of the wave packets (closely connected to the notion of well-localized particles) when the local wavelength of the modes of interest at the position of the Gaussian bump may be comparable to or longer than the width of the bump, potentially calling the applicability of~\eqref{eq:EER_def}-\eqref{eq:totEER_def} into question. However, the observation that the black-hole event horizon corresponds to a thermal system at the temperature~\eqref{eq:T_H}, is a robust and generic result, obtained independently via various methods which do not rely on wave packets, see e.g.~\cite{Gibbons:1976ue,Unruh:1976db,Hartle:1976tp,Gibbons:1976es,Gibbons:1977mu,Parikh:1999mf}. Furthermore, it should be emphasized that the GFs, are not tied to Hawking radiation. They correspond to frequency-dependent transmission coefficients for the~\emph{independent} scattering problem of~\emph{monochromatic} radiation impinging on the (in principle arbitrary) gravitational barrier of the black hole. The latter, acts as a frequency filter, and the resulting GFs encapsulate the intricacies of the potential in their frequency profile, subject only to the precision of the integration method. When the GFs are used in the context of Hawking radiation, their role is to modify the outgoing black-body radiation generated at the event horizon into what a distant stationary observer measures as the Hawking power spectrum. Thus, under the working assumption of negligible backreaction to the metric, the various features of the effective potential—such as bumps that we consider here—do not call into question the validity of the aforementioned Hawking formulas.

For the remainder of the article, we consider the Schwarzschild black hole of mass $M$ for our gravitational background, where
\beq
f(r)=1-\frac{2M}{r}=1-\frac{r_0}{r}\,,
\eeq
and we work in units of $r_0=1$.

\section{Near-horizon deformations}
\label{Sec:Near_horizon_deformations}

Near-horizon deformations are not on the same footing as environmental ones, as they are not constrained by current observations—neither in the gravitational nor in the electromagnetic spectrum. Indeed, most observational constraints primarily affect the geometry near the peak of the effective potential or around the innermost stable circular orbit.

From here onward, we distinguish between two types of bumps: those that contribute positively and those that contribute negatively. According to the simplest environmental model involving a modified mass function \cite{Konoplya:2018yrp}, positive bumps may be associated with phantom matter, while negative bumps correspond to ordinary matter. This interpretation follows from the observation that normal matter surrounding a black hole adds to the gravitational attraction, thereby lowering the effective potential barrier. However, this reasoning should be treated with caution: small amounts of phantom matter distributed over large space may not be sufficient to generate a positive bump, and conversely, ordinary matter could in some configurations lead to a negative contribution to the mass function.

\begin{figure*}[h!]
\begin{center}
\includegraphics[width=0.32\linewidth]{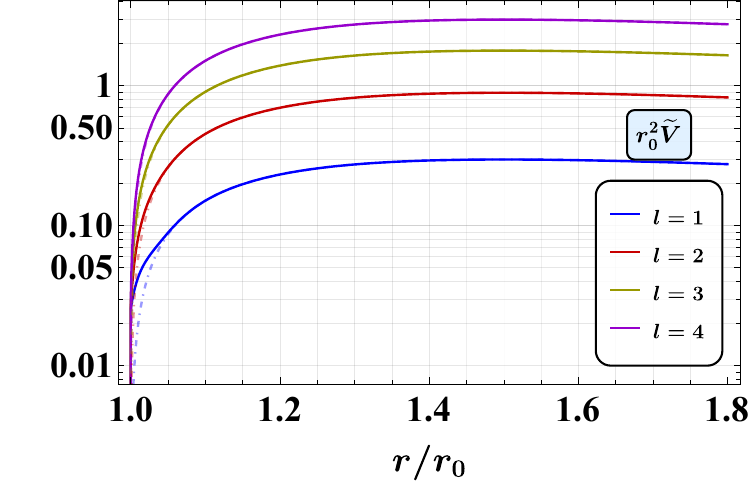}
\includegraphics[width=0.32\linewidth]{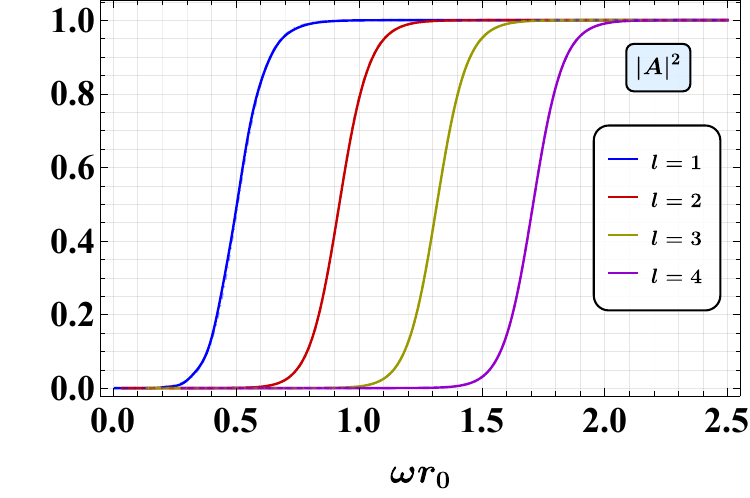}
\includegraphics[width=0.34\linewidth]{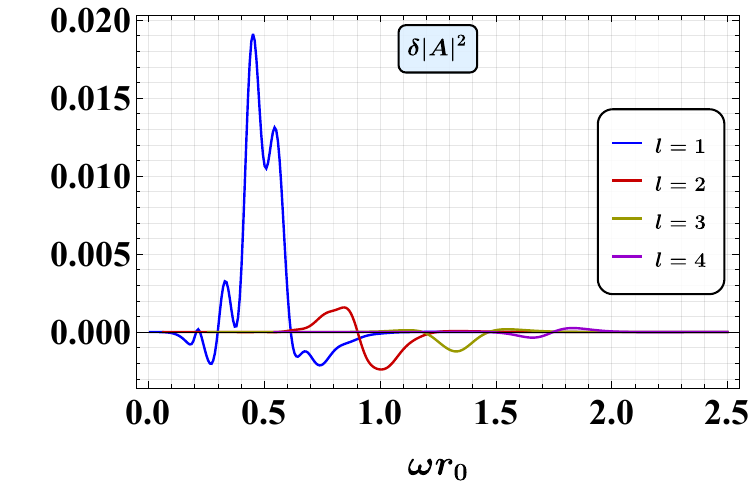}
\includegraphics[width=0.32\linewidth]{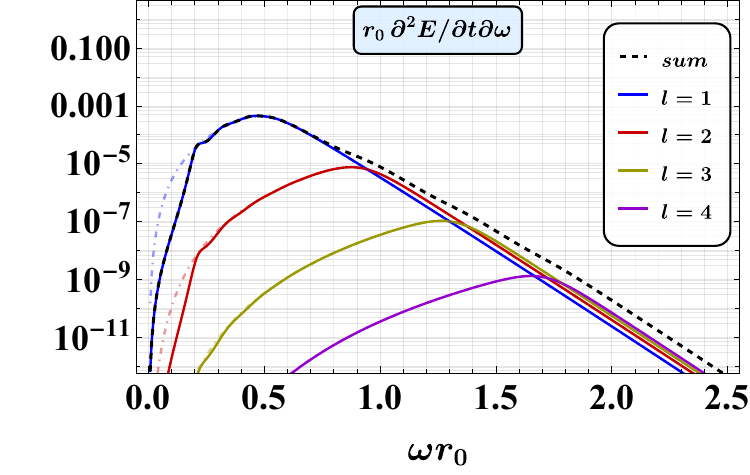}
\includegraphics[width=0.32\linewidth]{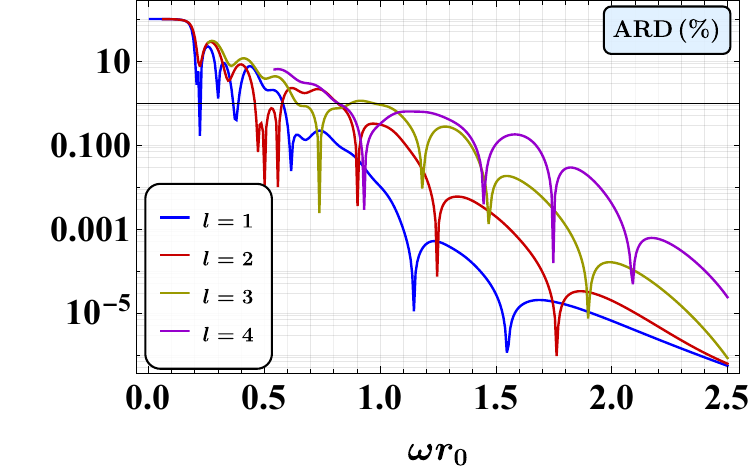}
\includegraphics[width=0.34\linewidth]{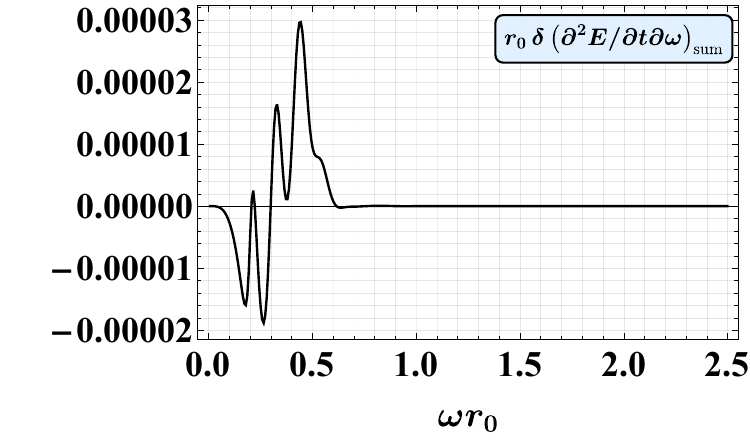}
\caption{\emph{Top left:} Deformation of the electromagnetic effective potential~\eqref{eq:Veff_em} by the Gaussian bump~\eqref{eq:Gaussian_bump_eps} with parameters $(\alp,r_m,\kappa)=(0.025,1.001,0.001)$, for the first 4 multipole numbers (colored solid curves). In all panels, dashed curves of the same color correspond to the no-bump limits.~\emph{Top middle:} Grey-body factors (GFs).~\emph{Top right}: Difference between the GFs and their no-bump limits.~\emph{Bottom left:} Differential energy-emission rates (EERs), along with their sum (black-dashed curve).~\emph{Bottom middle:} Percentage of absolute relative difference of the EERs from their no-bump limits.~\emph{Bottom right:} Difference between the sum of EERs and their no-bump limit.~The total emissivity of the black hole in the electromagnetic channel is enhanced by $\sim 1.1294 \%$.}
\label{fig:[A0025_rm1001_k0001]}
\end{center}
\end{figure*}

In Fig.~\ref{fig:[A0025_rm1001_k0001]}, we observe that relatively small positive bumps in the near-horizon region of the effective potential for electromagnetic perturbations are barely visible for the $\ell=1$ case and become entirely negligible for higher multipoles.
The corresponding grey-body factors deviate from the Schwarzschild values by only a few percent for $\ell=1$, and by a fraction of a percent for $\ell \geq 2$. This results in a small shift in the grey-body factor for $\ell=1$, and an almost negligible shift for higher $\ell$. Consequently, the differential energy emission rate (per unit frequency), even though it gets strongly modified (at the order of tens of percent) at very low frequencies~$\omega r_0\ll1$, it only exhibits a mild modification (around one percent) for the frequency range corresponding to its peak. Furthermore, notice that the emission at small frequencies is suppressed and at larger frequencies is enhanced. As a result of the above, the total energy emission rate changes insignificantly, with the total emission enhanced by approximately one percent.

\begin{figure*}[h!]
\begin{center}
\includegraphics[width=0.32\linewidth]{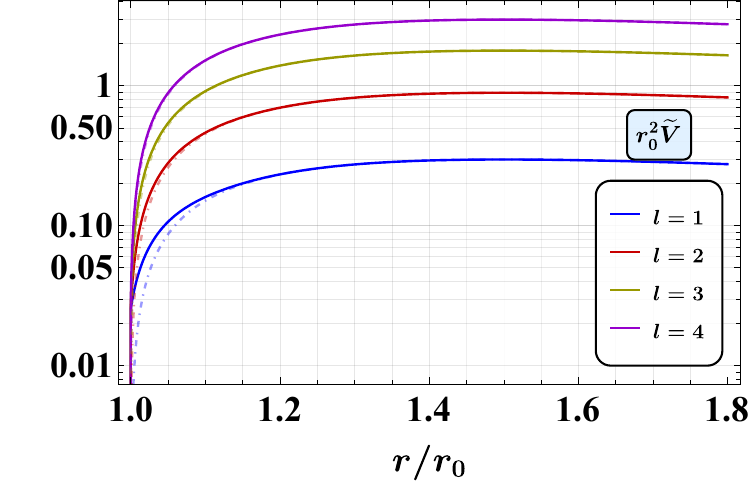}
\includegraphics[width=0.32\linewidth]{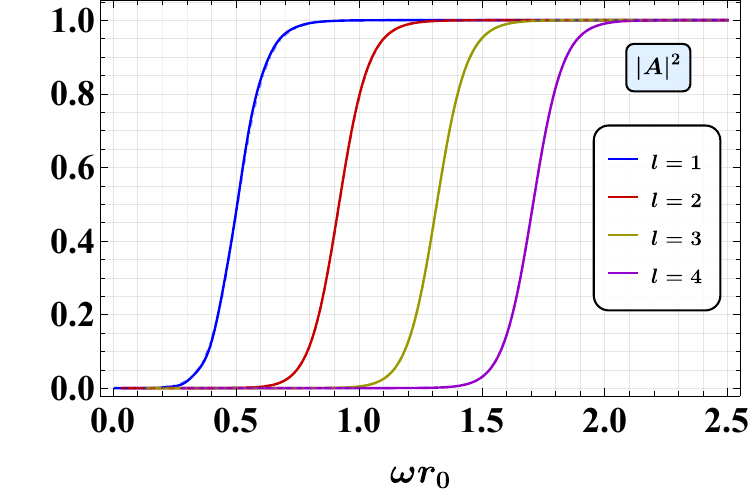}
\includegraphics[width=0.34\linewidth]{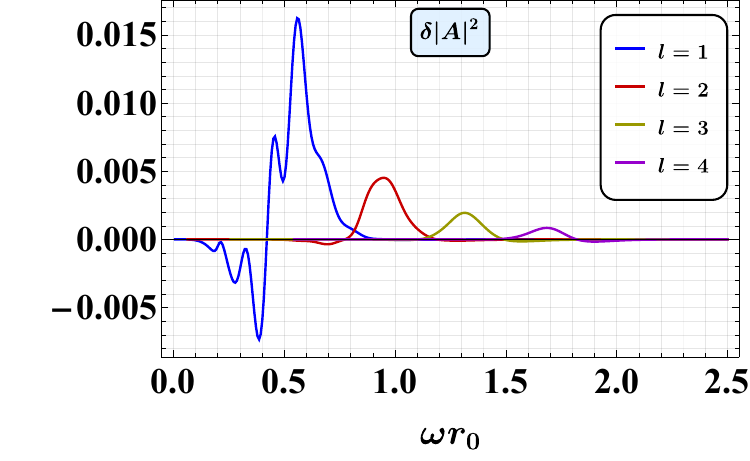}
\includegraphics[width=0.32\linewidth]{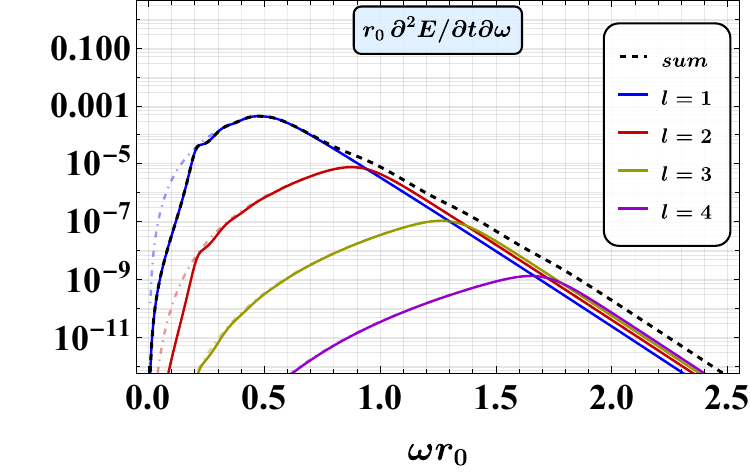}
\includegraphics[width=0.32\linewidth]{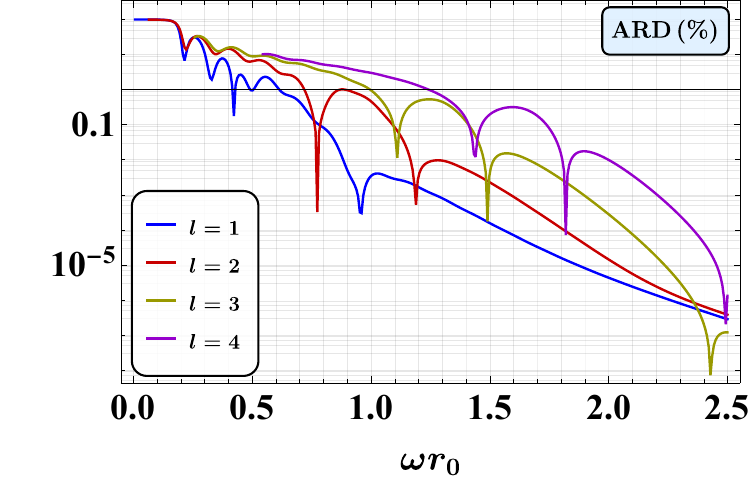}
\includegraphics[width=0.34\linewidth]{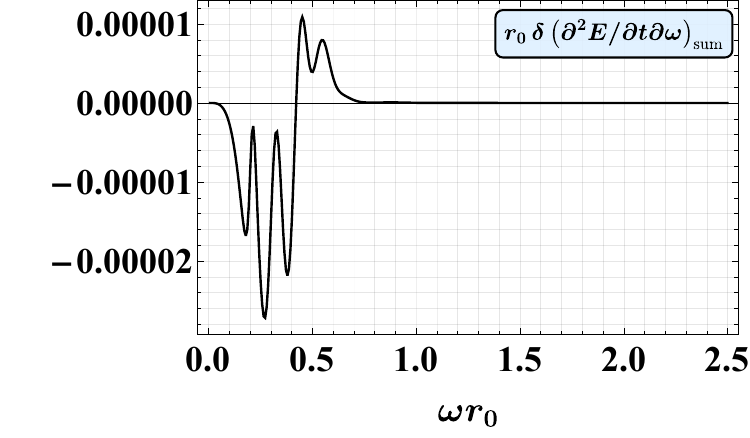}
\caption{When the effective potential~\eqref{eq:Veff_em} is deformed by a Gaussian bump~\eqref{eq:Gaussian_bump_eps} with parameters $(\alp,r_m,\kappa)=(0.025,1.001,0.01)$, the total emissivity of the black hole in the electromagnetic channel is reduced by $\sim 2.1848 \%$. For details on the contents of panels, see caption of Fig.~\ref{fig:[A0025_rm1001_k0001]}.}
\label{fig:[A0025_rm1001_k001]}
\end{center}
\end{figure*}

In Fig.~\ref{fig:[A0025_rm1001_k001]}, we observe that changing only a single parameter of the bump — specifically, its width as determined by $\kappa$ — can lead to an effect opposite to that seen previously.
Although still small, this modification results in a suppression of the total radiation by approximately two percent. This behavior can be hypothetically explained by noting that a larger width corresponds to a less localized deformation of the potential, which tends to affect the lower-frequency modes more significantly than the higher ones.

\begin{figure*}[h!]
\begin{center}
\includegraphics[width=0.32\linewidth]{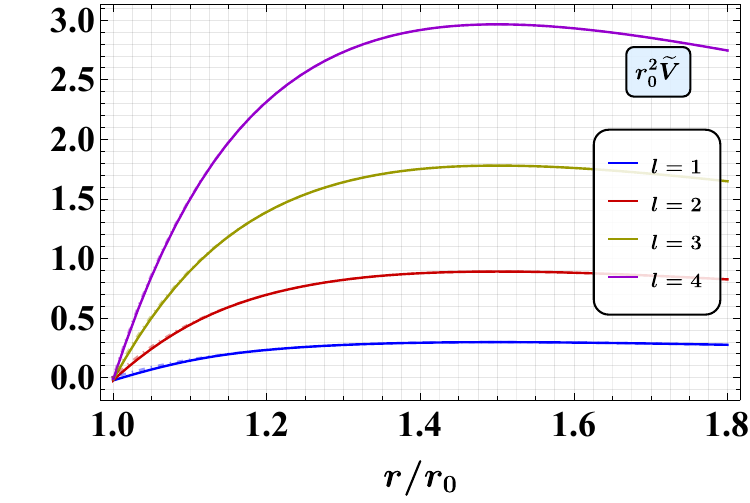}
\includegraphics[width=0.32\linewidth]{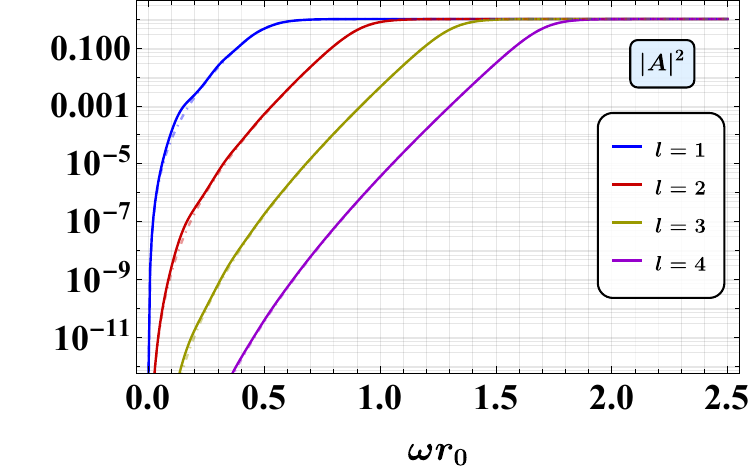}
\includegraphics[width=0.34\linewidth]{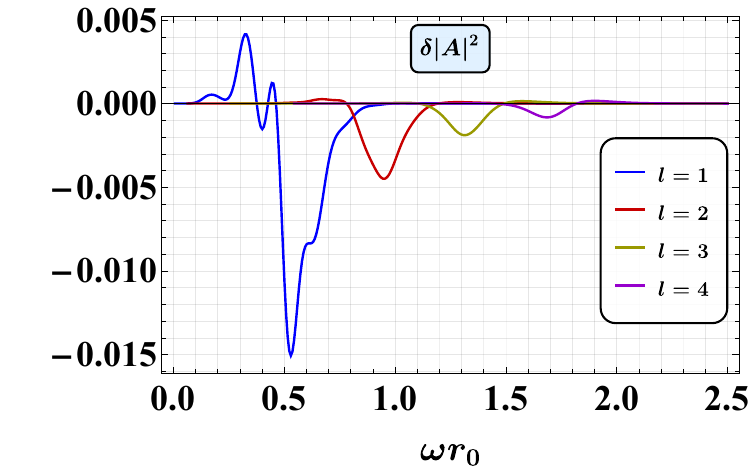}
\includegraphics[width=0.32\linewidth]{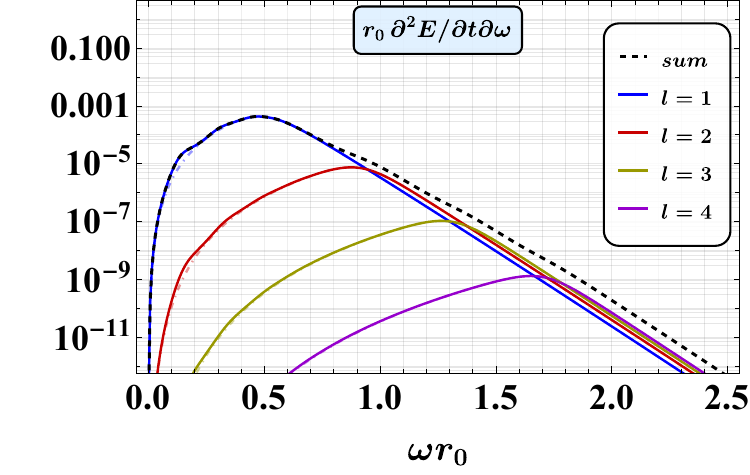}
\includegraphics[width=0.32\linewidth]{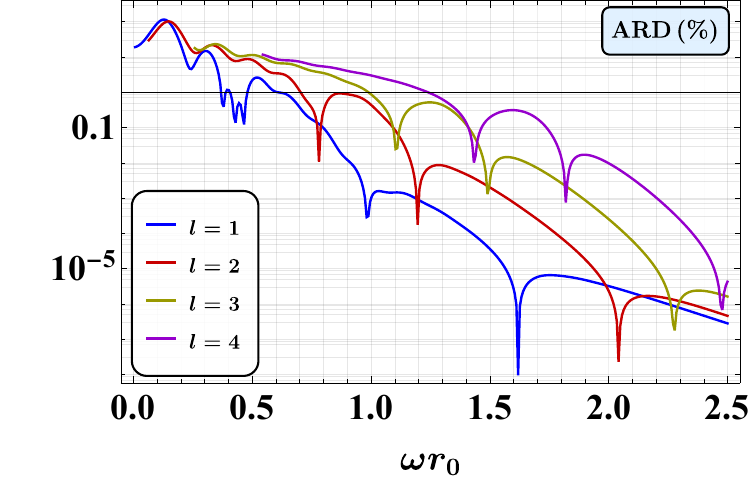}
\includegraphics[width=0.34\linewidth]{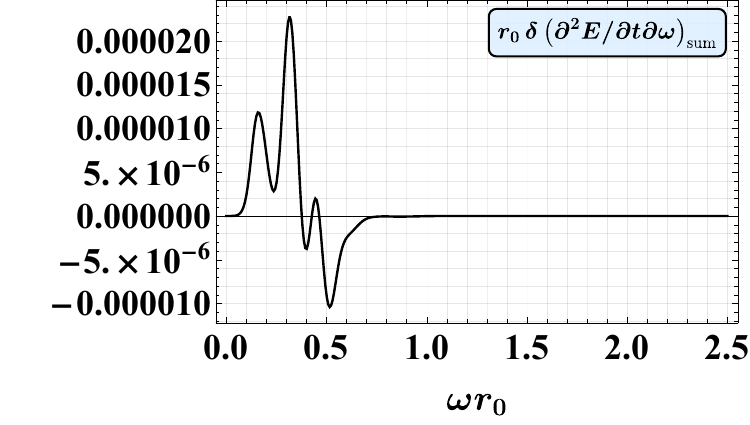}
\caption{When the effective potential~\eqref{eq:Veff_em} is deformed by a Gaussian dip~\eqref{eq:Gaussian_bump_eps} with parameters $(\alp,r_m,\kappa)=(-0.025,1.001,0.01)$, the total emissivity of the black hole in the electromagnetic channel is enhanced by $\sim 1.2899 \%$. For details on the contents of the panels, see caption of Fig.~\ref{fig:[A0025_rm1001_k0001]}.}
\label{fig:[A-0025_rm1001_k001]}
\end{center}
\end{figure*}

When the negative bump is replaced by a positive one, as shown in Fig.~\ref{fig:[A-0025_rm1001_k001]}, the effect remains of the same order of magnitude; however, in this case, the emission at lower frequencies is enhanced, while the emission at higher frequencies is slightly suppressed.
As before, a broader bump affects the lower-frequency modes more strongly, resulting in an overall increase in the total energy emission rate by approximately one percent.

It should also be noted that in all of the above examples, the Hawking temperature of the black hole remains unchanged when the bump is added to the spacetime. Therefore, the observed shifts in the energy emission rates are entirely due to modifications in the grey-body factors. It is evident that if the near-horizon deformation were to alter the Hawking temperature, the emission rates would be affected much more significantly, since the temperature appears in the exponent of the Boltzmann factor in the Hawking radiation formula.

The most important observation regarding near-horizon deformations is that when the height of the bump is not negligibly small — though still apparently smaller than the main potential peak — the impact on grey-body factors at low frequencies, and consequently on the energy emission rate, can be significant, reaching tens of percent for the emission rates when $\alp r_0^2$ is comparable to the main potential peak. This effect is illustrated in Table~\ref{tab:1} for the case $|\alpha r_{0}^2| = 0.025$, corresponding to about $10\%$ of the main peaks of effective potentials for the dominant modes i.e. for the lowest multipole numbers $l$.\footnote{Since the main peak of the potentials~\eqref{eq:Veff_em} and~\eqref{eq:Veff_Dirac} increases with $l$, it follows that the same amplitude for the deformations corresponds to a smaller percentage of the main peak as the multipole number increases.} In the next section, we will demonstrate that such a substantial influence does not occur for environmental deformations located in the far zone.

\section{Deformations due to environment}
\label{Sec:Far_zone_deformations}

Unlike near-horizon deformations, environmental deformations are subject to several constraints. First, at sufficiently large distances from the black hole, the spacetime must reproduce the well-tested post-Newtonian behavior. Moreover, typical astrophysical environments of black holes—such as accretion disks—are known to be much less massive than the black hole itself, contributing at most about $10^{-6}$ to $10^{-7}$ of the black hole mass, and are spread over spatial regions that exceed the black hole's size by several orders of magnitude \cite{Abramowicz:2011xu}. Such environments are expected to produce negligible effects on grey-body factors and Hawking radiation, since the resulting bumps would have extremely small amplitude (at least 5-6 orders of magnitude smaller than the main peak) and very large $\kappa$. In order to observe any noticeable deviation due to bumps in the far zone, one would need to significantly increase these parameters to, in our view, unrealistically large values — corresponding to some highly energetic and non-linear processes occurring in the vicinity of the black hole.
   
In realistic astrophysical scenarios, accretion disks around black holes are typically thin and extend from the innermost stable circular orbit (ISCO) outward to several orders of magnitude in radius, depending on the accretion environment. For Schwarzschild black holes, the ISCO lies at $r = 6M$, while for rapidly rotating Kerr black holes it can approach $r = M$ for prograde orbits. Although the disk may formally extend to $r \sim 10^5 M$ or more in some systems, such as X-ray binaries or active galactic nuclei, most of the disk mass and luminosity is concentrated relatively close to the black hole. For standard thin disk models, the surface density typically peaks at a few tens of $M$, so that the effective center of mass of the disk lies around $r \sim 10$–$50M$ \cite{Novikov:1973kta,Page:1974he}. As a result, gravitational or radiative influence from the disk is expected to be most significant in this region, and any deviations in the near-horizon geometry due to disk matter are negligible unless the disk is extremely compact or massive. Thus, for a spherically symmetric solution, choosing the bump’s center at $r_{m} = 10$ is a reasonable — if not somewhat exaggerated — illustration of environmental effects.
 
\begin{figure*}[h!]
\begin{center}
\includegraphics[width=0.32\linewidth]{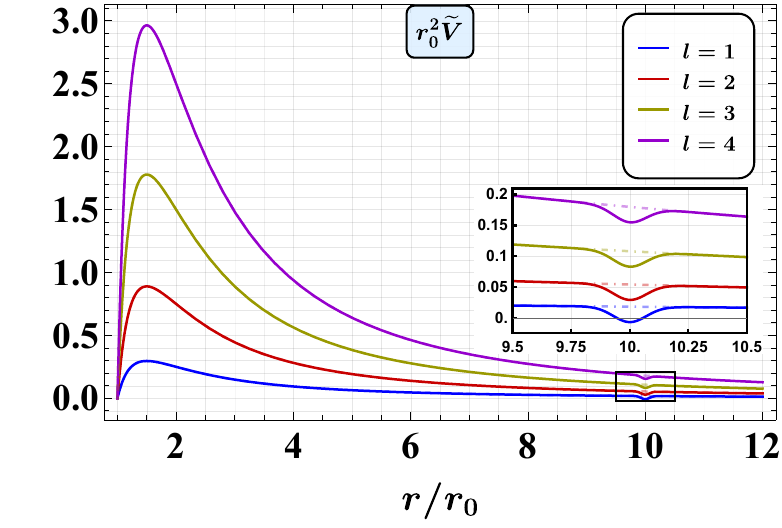}
\includegraphics[width=0.32\linewidth]{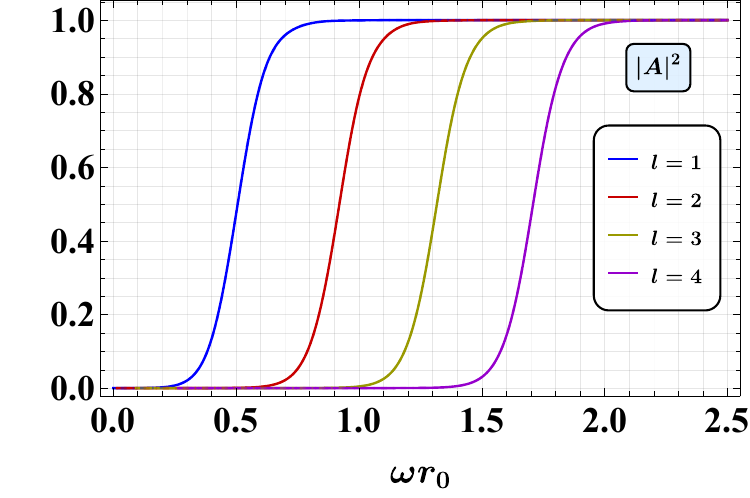}
\includegraphics[width=0.34\linewidth]{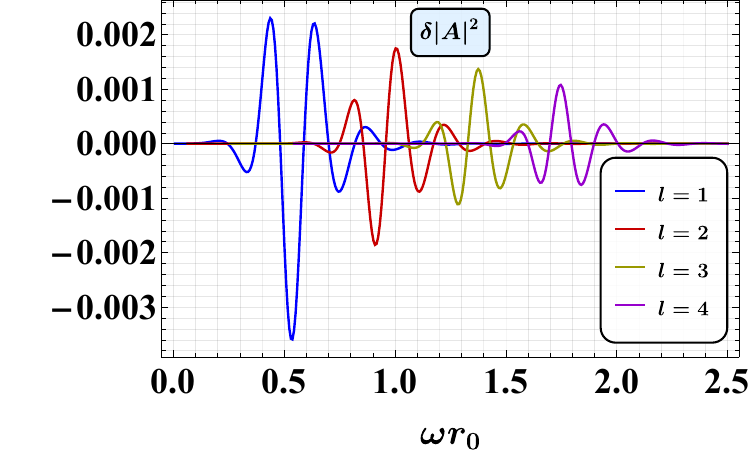}
\includegraphics[width=0.32\linewidth]{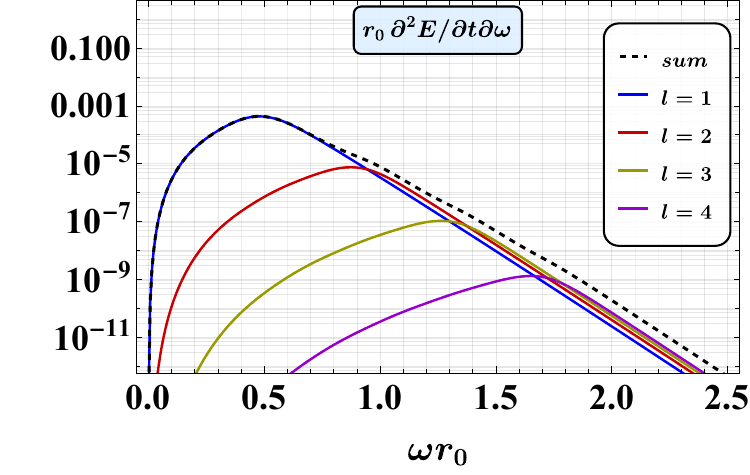}
\includegraphics[width=0.32\linewidth]{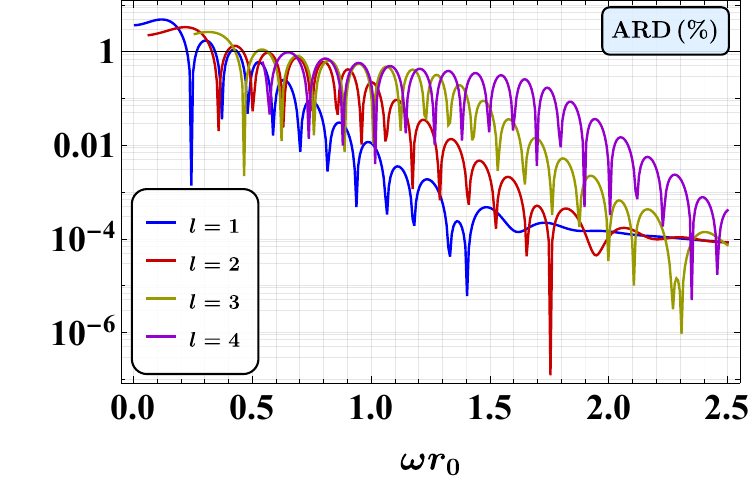}
\includegraphics[width=0.34\linewidth]{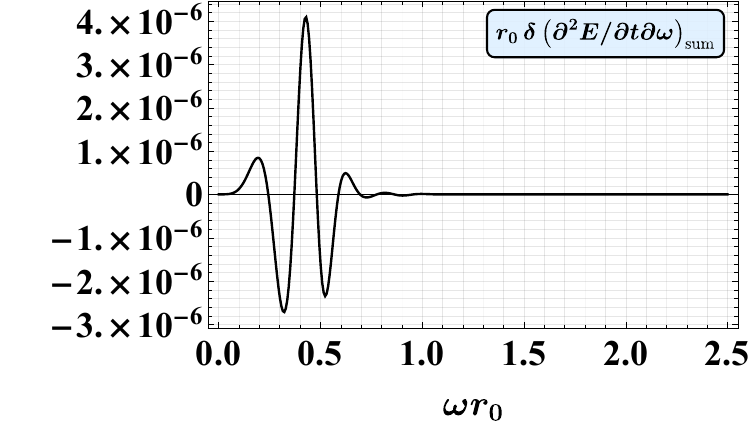}
\caption{When the effective potential~\eqref{eq:Veff_em} is deformed by a Gaussian bump~\eqref{eq:Gaussian_bump_eps} with parameters $(\alp,r_m,\kappa)=(-0.025,10,0.01)$, the total emissivity of the black hole in the electromagnetic channel is enhanced by $\sim 0.01679 \%$. For details on the contents of the panels, see caption of Fig.~\ref{fig:[A0025_rm1001_k0001]}. The inlaid plot in the top-left panel provides a magnified view of the region indicated by the black rectangle in the main plot.}
\label{fig:[A-0025_rm10_k001]}
\end{center}
\end{figure*}

In Fig.~\ref{fig:[A-0025_rm10_k001]}, we observe that in the bosonic sector, a bump with the same depth, $\alpha = -0.025$, as in the previous near-horizon example — but now located in the far zone — leads to an enhancement in the energy emission rate of only about $\sim 0.01679\%$. This is two orders of magnitude smaller than the corresponding effect in the near-horizon case.

\begin{figure*}[h!]
\begin{center}
\includegraphics[width=0.32\linewidth]{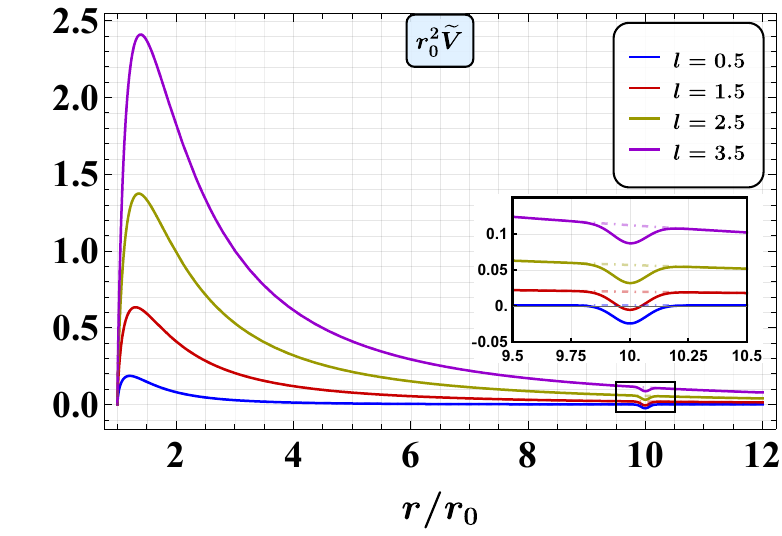}
\includegraphics[width=0.32\linewidth]{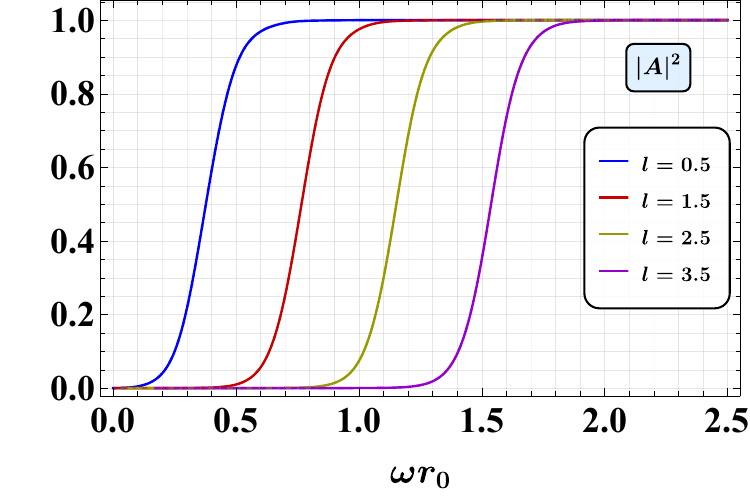}
\includegraphics[width=0.34\linewidth]{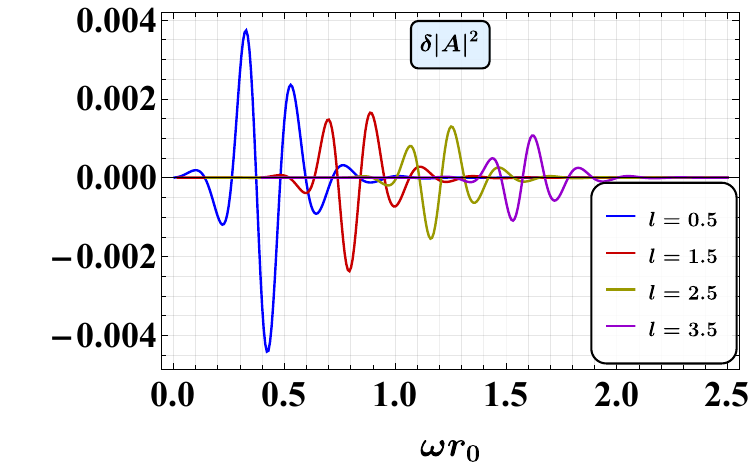}
\includegraphics[width=0.32\linewidth]{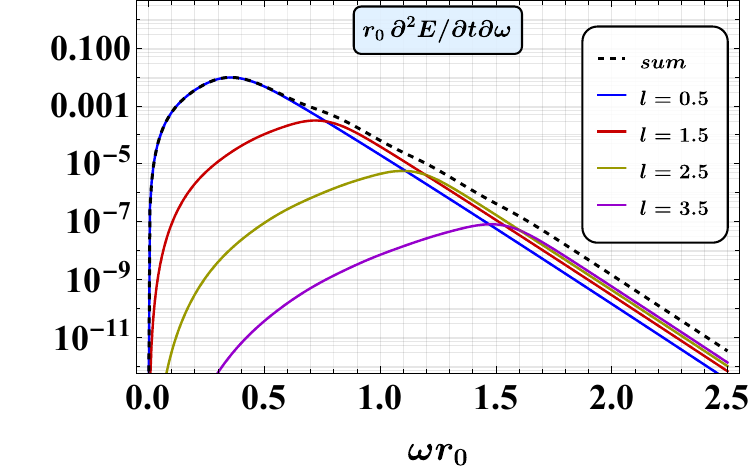}
\includegraphics[width=0.32\linewidth]{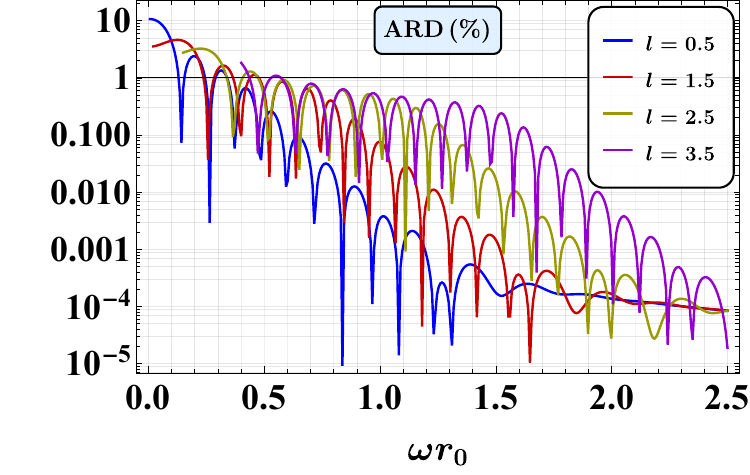}
\includegraphics[width=0.34\linewidth]{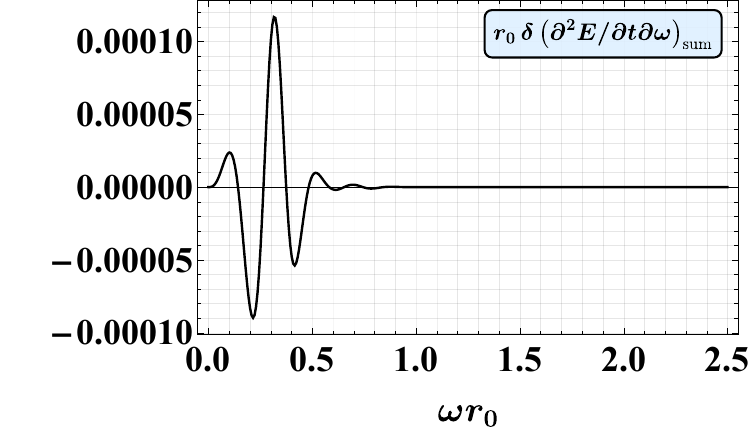}
\caption{When the effective potential~\eqref{eq:Veff_Dirac} is deformed by a Gaussian dip~\eqref{eq:Gaussian_bump_eps} with parameters $(\alp,r_m,\kappa)=(-0.025,10,0.01)$, the total emissivity of the black hole in the Dirac channel is reduced by $\sim 0.00356 \%$. For details on the contents of the panels, see caption of Fig.~\ref{fig:[A0025_rm1001_k0001]}. The inlaid plot in the top-left panel provides a magnified view of the region indicated by the black rectangle in the main plot.}
\label{fig:[A-0025_rm10_k001][Dirac]}
\end{center}
\end{figure*}

In the fermionic sector,~Fig.~\ref{fig:[A-0025_rm10_k001][Dirac]}, we see a slight suppression of the emission, amounting to approximately $\sim 0.00356\%$. These results seem to suggest that, for relatively small environmental deformations of the geometry, fermions respond oppositely to bosons: fermionic emission is suppressed for normal matter (negative bumps) and enhanced for phantom matter (positive bumps), whereas bosonic emission is enhanced by normal bumps and suppressed by phantom ones. Nevertheless, as our analysis in Sec.~\ref{Sec:Variations_of_location} reveals, the exact location of the deformation affects the response of the two sectors to the deformation, resulting to various scenarios such as e.g. suppression of the emission in both channels.

At the same time, we clearly see that the impact of environmental bumps is two or more orders of magnitude smaller than that of near-horizon deformations of the same height or depth. For instance, according to the data presented in table~\ref{tab:1}, for $\alpha = -0.025$ (corresponding to a depth of the deformation of about $10\%$ of the main peak of the dominant-mode effective potentials), the effect in the far zone is a reduction of only $-0.0006966\,\%$ for fermionic emission, whereas the same bump located near the horizon produces a significantly larger effect with an enhancement of $7.74883\,\%$. Note that the depth $\alpha = -0.025$ is extraordinarily large — several orders of magnitude greater than what would be expected from the mass of a typical accretion disk.

\section{General laws for the total emissivity}
\label{Sec:General_laws}

While deriving a general law for the variation of the energy emission rate with respect to the bump parameters $\alpha$, $r_{m}$, and $\kappa$ is an unfeasible task, we can nevertheless observe the behavior in two limiting regimes: when the height and width of the bump are small, and when the bump is located far from the black hole, i.e., at large $r_m$.
  
\subsection{Variations of the height of the deformation}
\label{Sec:Variations_of_height}

In order to study the impact of the height of the deformation~\eqref{eq:Gaussian_bump_eps}, we vary the values of the height parameter $\alp\,r_0^2$ while keeping the width $\kap\,r_0^{-2}$ and location $r_m/r_0$ parameters fixed. We consider values for the height parameter that are not too large in comparison with the main peak of the effective potentials for the dominant modes of each type of perturbation which are $V_0\simeq 0.296\,r_0^{-2}$ for electromagnetic and $V_0\simeq 0.187\,r_0^{-2}$ for Dirac test fields. In particular, the maximum considered value of $|\alp\,r_0^2|=0.025$, corresponds to approximately $10\%$ of $V_0\,r_0^2$. Even though for near-horizon deformations that are sufficiently well localized near $r_0$, such deformation amplitudes may be realistic, in the far-zone, on the contrary, they are several orders of magnitude larger than typical environmental effects. Nevertheless, such extreme values will also be considered in our far-zone analysis for illustrative purposes.

According to the data presented in table~\ref{tab:1}, we find that near-horizon deformations that are highly localized (i.e., for small values of $\kappa$), and near the event horizon, suppress the emission of bosons for negative bumps and enhance it for positive bumps. On the other hand, fermionic emission responds to the same deformations in the opposite way. The combined total energy emission rate of both channels follows closely the fermionic sector since the black hole emits fermions with about one order of magnitude more power compared to bosons.

\begin{table}
\centering
\resizebox{\textwidth}{!}{
\begin{tabular}{|c||c|c||c|c||c|c||}
\cline{2-7}
\multicolumn{1}{c||}{} & \multicolumn{2}{c||}{EM} & \multicolumn{2}{c||}{Dirac} & \multicolumn{2}{c||}{EM $+$ Dirac} \\
\hline\hline
\multicolumn{1}{|c||}{$\alp\,r_0^2$} & \multicolumn{1}{c|}{$r_m=1.001\,r_0$} & \multicolumn{1}{c||}{$r_m=10\,r_0$} & \multicolumn{1}{c|}{$r_m=1.001\,r_0$} & \multicolumn{1}{c||}{$r_m=10\,r_0$} & \multicolumn{1}{c|}{$r_m=1.001\,r_0$} & \multicolumn{1}{c||}{$r_m=10\,r_0$} \\
\hline
\hline
-0.025 & -1.5826~$\%$ & 0.00549353~$\%$ & 7.74883~$\%$ & -0.000696606~$\%$ & 7.34125~$\%$ & -0.000426229~$\%$ \\
-0.015 & -0.952475~$\%$ & 0.00331653~$\%$ & 4.49942~$\%$ & -0.00037111~$\%$ & 4.26129~$\%$ & -0.000210039~$\%$ \\
-0.01 & -0.638313~$\%$ & 0.00221782~$\%$ & 3.12476~$\%$ & -0.000231805~$\%$ & 2.96039~$\%$ & -0.000124808~$\%$ \\
-0.0075 & -0.475702~$\%$ & 0.00166592~$\%$ & 2.54018~$\%$ & -0.000168006~$\%$ & 2.40845~$\%$ & -0.0000879025~$\%$ \\
-0.005 & -0.310018~$\%$ & 0.00111231~$\%$ & 1.84422~$\%$ & -0.000108107~$\%$ & 1.75013~$\%$ & -0.0000548004~$\%$ \\
-0.0025 & -0.150837~$\%$ & 0.000557008~$\%$ & 0.959715~$\%$ & -0.0000521055~$\%$ & 0.911208~$\%$ & -0.0000255002~$\%$ \\
-0.001 & -0.0593786~$\%$ & 0.000223007~$\%$ & 0.39045~$\%$ & -0.0000203748~$\%$ & 0.370802~$\%$ & -9.74419$\times 10^{-6}$~$\%$ \\
-0.00075 & -0.0444164~$\%$ & 0.000167281~$\%$ & 0.293569~$\%$ & -0.0000152227~$\%$ & 0.278806~$\%$ & -7.25116$\times 10^{-6}$~$\%$ \\
-0.0005 & -0.0295329~$\%$ & 0.000111538~$\%$ & 0.196185~$\%$ & -0.0000101095~$\%$ & 0.186326~$\%$ & -4.79612$\times 10^{-6}$~$\%$ \\
-0.00025 & -0.0147276~$\%$ & 0.0000557774~$\%$ & 0.0983219~$\%$ & -5.03529$\times 10^{-6}$~$\%$ & 0.093384~$\%$ & -2.37908$\times 10^{-6}$~$\%$ \\
-0.0001 & -0.00588174~$\%$ & 0.000022313~$\%$ & 0.0393825~$\%$ & -2.00944$\times 10^{-6}$~$\%$ & 0.0374054~$\%$ & -9.4707$\times 10^{-7}$~$\%$ \\
\hline
0.0001 & 0.00586939~$\%$ & -0.0000223157~$\%$ & -0.0394527~$\%$ & 2.00322$\times 10^{-6}$~$\%$ & -0.0374731~$\%$ & 9.40999$\times 10^{-7}$~$\%$ \\
0.00025 & 0.0146504~$\%$ & -0.0000557944~$\%$ & -0.0987609~$\%$ & 4.99635$\times 10^{-6}$~$\%$ & -0.0938073~$\%$ & 2.34109$\times 10^{-6}$~$\%$ \\
0.0005 & 0.029224~$\%$ & -0.000111606~$\%$ & -0.197943~$\%$ & 9.95376$\times 10^{-6}$~$\%$ & -0.18802~$\%$ & 4.6442$\times 10^{-6}$~$\%$ \\
0.00075 & 0.0437213~$\%$ & -0.000167434~$\%$ & -0.297527~$\%$ & 0.0000148722~$\%$ & -0.282622~$\%$ & 6.90933$\times 10^{-6}$~$\%$ \\
0.001 & 0.0581428~$\%$ & -0.00022328~$\%$ & -0.397499~$\%$ & 0.0000197518~$\%$ & -0.377597~$\%$ & 9.13649$\times 10^{-6}$~$\%$ \\
0.0025 & 0.143102~$\%$ & -0.00055871~$\%$ & -1.00464~$\%$ & 0.0000482116~$\%$ & -0.954511~$\%$ & 0.000021702~$\%$ \\
0.005 & 0.278883~$\%$ & -0.00111912~$\%$ & -2.03913~$\%$ & 0.0000925312~$\%$ & -1.93788~$\%$ & 0.0000396078~$\%$ \\
0.0075 & 0.407659~$\%$ & -0.00168124~$\%$ & -3.09429~$\%$ & 0.000132961~$\%$ & -2.94133~$\%$ & 0.000053719~$\%$ \\
0.01 & 0.529681~$\%$ & -0.00224505~$\%$ & -4.16402~$\%$ & 0.000169502~$\%$ & -3.959~$\%$ & 0.0000640377~$\%$ \\
0.015 & 0.75425~$\%$ & -0.00337779~$\%$ & -6.33085~$\%$ & 0.000230929~$\%$ & -6.02138~$\%$ & 0.0000733048~$\%$ \\
0.025 & 1.12943~$\%$ & -0.00566371~$\%$ & -10.7067~$\%$ & 0.00030721~$\%$ & -10.1897~$\%$ & 0.0000464085~$\%$ \\
\hline\hline
\end{tabular}
}
\caption{Relative difference between the total emissivity in the presence of various localized deformations~\eqref{eq:Gaussian_bump_eps} of the effective potential with $\kap=0.001\,r_0^2$, and the undeformed limits of $r_0^2\,dE/dt\simeq 0.000134554185$, and $r_0^2\,\dd E/\dd t \simeq 0.002945989552$ for the EM and Dirac channels respectively. Positive (negative) values indicate enhancement (suppression) of the emission due to the deformation. The contributions from the the first four multipole numbers have been taken into account. See also Fig.~\ref{fig:TotalEER_RD_alp}.}
\label{tab:1}
\end{table}

\begin{figure*}[h!]
\begin{center}
\includegraphics[width=0.48\linewidth]{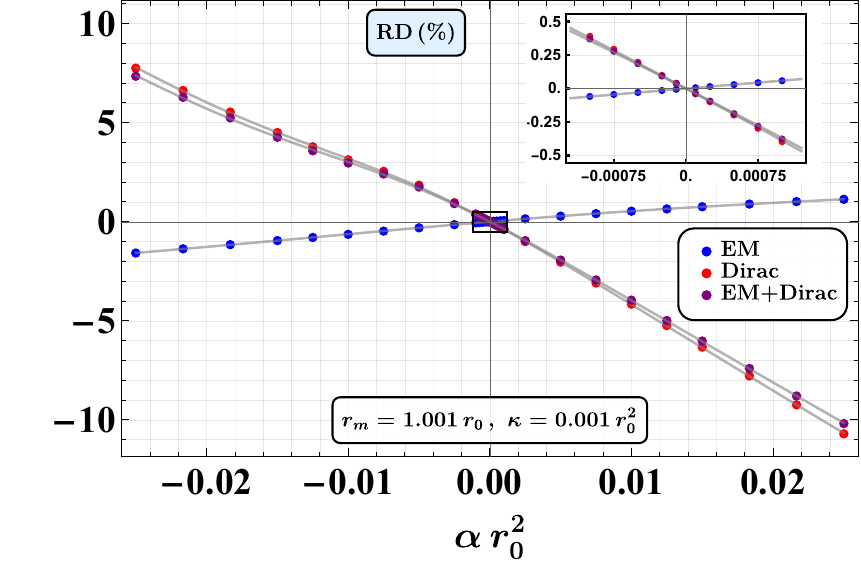}
\includegraphics[width=0.50\linewidth]{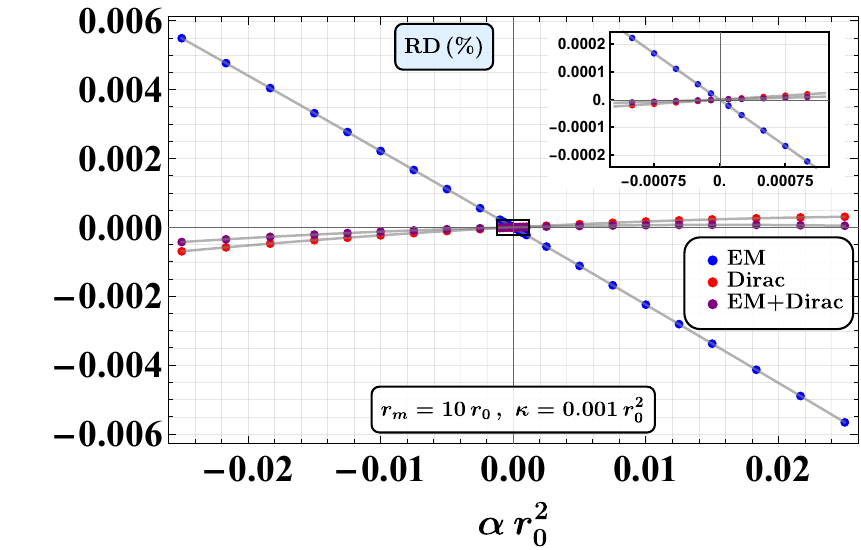}
\caption{Relative difference between the total emissivity $r_0^2 \dd E/ \dd t$ and its Schwarzschild limit, in the case of localized deformations~\eqref{eq:Gaussian_bump_eps} near the event horizon (left panels), and in the far zone (right panels). The extremal values of $\alp\,r_0^2=\pm0.025$ correspond to approximately $10\%$ of the main peak heights. The gray curves correspond to the fits of the numerical data as given in Eqs.~\eqref{eq:Fit_alp_EM_1001}-\eqref{eq:Fit_alp_emDir_10}. The inlaid plots provide a magnified view of the region near the origin, as indicated by the black rectangles in the main plots.}
\label{fig:TotalEER_RD_alp}
\end{center}
\end{figure*}

We further observe that as a general trend for both near-horizon and far-zone deformations, the total emissivity exhibits a linear dependence on the height of the deformations when $\alp\, r_0^2$ is sufficiently small (of the order of~$0.1\%$), and in some cases, this linearity extends even for large values of $\alp\, r_0^2$, of the order of $10\%$ of the main peaks of the effective potentials, see~Fig.~\ref{fig:TotalEER_RD_alp}.

\begin{figure*}[h!]
\begin{center}
\includegraphics[width=0.49\linewidth]{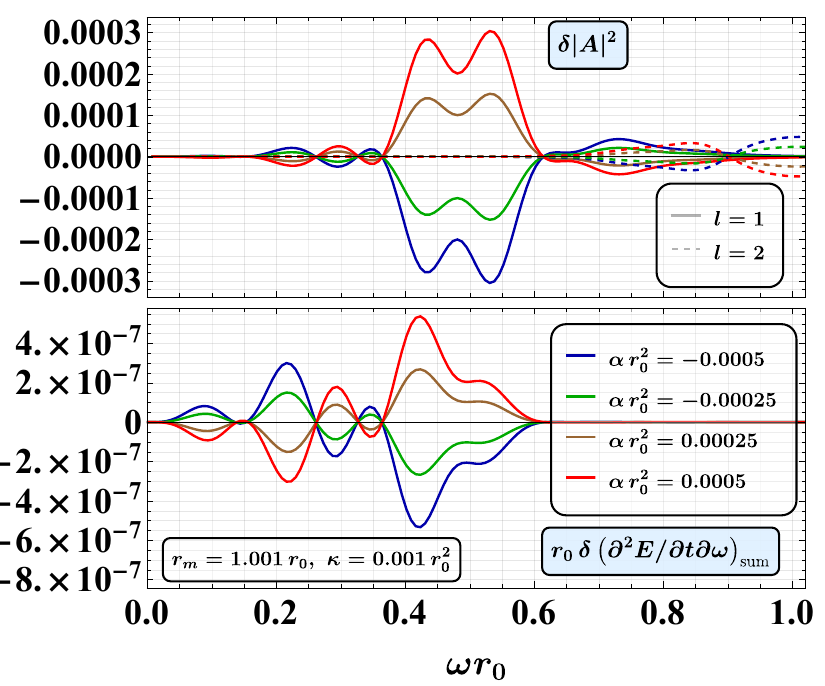}
\includegraphics[width=0.49\linewidth]{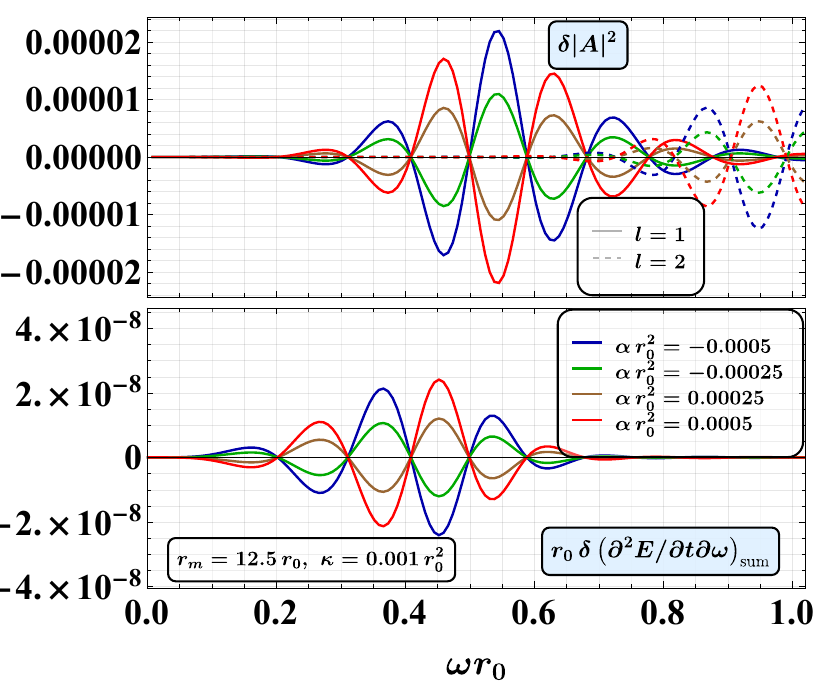}
\caption{The effect of the amplitude of the deformation~\eqref{eq:Gaussian_bump_eps} for the bosonic channel of emission. Top panels: The difference between the GFs in the presence of deformations and their Schwarzschild limit, for the first (solid curves) and second (dashed curves) multipole number. Bottom panels: The corresponding difference in the sum of the energy emission rates for the first four multipole numbers. Left and right panels correspond to near-horizon and far-zone deformations respectively.}
\label{fig:DeltaGFandEER_alpha_variations}
\end{center}
\end{figure*}

To further understand the linear dependence on $\alp\,r_0^2$, in Fig.~\ref{fig:DeltaGFandEER_alpha_variations}, we present the changes in the grey-body factors and the corresponding changes on the differential energy emission rate for small-amplitude deformations. We consider the bosonic sector for our illustration but similar conclusions can be drawn even for the emission of fermions. In the top panels of Fig.~\ref{fig:DeltaGFandEER_alpha_variations}, one can see that a doubling of amplitude of the deformation, approximately doubles the change in the GFs and consequently to the EERs. Furthermore we see that the dominant-mode GFs exhibit the largest difference from the Schwarzschild values near the frequencies corresponding to the peak of the EERs for the bosonic emission for the Schwarzschild BH, that is around $\omega r_0 \simeq0.5$. The modifications of the GFs for the higher multipole numbers, are less prominent than those of the dominant mode and shifted to higher frequencies. As such, the contributions of their modifications to the total emission are severely subdominant.

An important observation is that the differences in the GFs exhibit an alternating pattern, changing between positive and negative values at various frequencies. Notice that this alternating frequency profile is very smooth and periodic for the far-zone deformations and significantly more irregular for near-horizon ones. These features are directly reflected on the frequency profiles of the EER differences as shown in the bottom panels of Fig.~\ref{fig:DeltaGFandEER_alpha_variations}, and help us understand why the effects of near-horizon deformations on the total emission are more pronounced than that of the far-zone ones. The near-horizon irregularities, (stemming perhaps from various kinds of interplay between the deformation and the main peak of the effective potential) typically tilt the balance clearly in favor of either enhancement or suppression of the total emission. On the other hand, for almost periodic alternating patterns observed for far-zone deformations in Fig.~\ref{fig:DeltaGFandEER_alpha_variations}, the contributions to the integral for the computation of the total emissivity cancel out almost entirely, resulting in significantly smaller modifications to the total emission.

Thus, we identify two reasons why the emission rates are much more strongly influenced by near-horizon deformations than by those in the far zone. The first reason is the significantly higher sensitivity of the grey-body factors to near-horizon deformations, which, in turn, arises from the proximity between the main potential peak and the center of the near-horizon deformation. The second reason is the nearly periodic dependence of the energy emission rate shift on frequency in the case of far-zone deformations, which causes the total shift in the energy emission to remain very small due to cancellations.

The fitting functions $F$ for the numerical data as presented in Fig.~\ref{fig:TotalEER_RD_alp} are given in Eqs.~\eqref{eq:Fit_alp_EM_1001}-\eqref{eq:Fit_alp_emDir_10}. The left superscript in $F$, denotes the channel of emission,~$\bfs{EM}$ for bosons,~$\bfs{Dir}$ for fermions, and~$\bfs{EM+Dir}$ for the combined total emission of both channels. The right superscript, denotes the value of $r_m/r_0$, while the subscript indicates the value of $\kap\,r_0^{-2}$. The values of $F(\alp\,r_0^2)$ correspond to the percentage of relative difference of the total emissivity for dimensionless height parameter $\alp\,r_0^2$ and the Schwarzschild limit.
\ba
^{\bfs{EM}}F^{1.001}_{0.001}\left(\alp r_0^2\right)&=& \frac{\left(\alp r_0^2\right) \left[59.513 -874\left(\alp r_0^2\right) +42817\left(\alp r_0^2\right)^2\right]}{1 -5.880\left(\alp r_0^2\right) +909\left(\alp r_0^2\right)^2}\,,\label{eq:Fit_alp_EM_1001}\\
^{\bfs{Dir}}F^{1.001}_{0.001}\left(\alp r_0^2\right)&=& \frac{\left(\alp r_0^2\right) \left[-383.908 -25513\left(\alp r_0^2\right) -1122769\left(\alp r_0^2\right)^2\right]}{1  +52.001\left(\alp r_0^2\right) +2779\left(\alp r_0^2\right)^2} \,,\label{eq:Fit_alp_Dir_1001}\\
^{\bfs{EM+Dir}}F^{1.001}_{0.001}\left(\alp r_0^2\right)&=& \frac{\left(\alp r_0^2\right) \left[-364.415 -24138\left(\alp r_0^2\right) -1058570\left(\alp r_0^2\right)^2 \right]}{1 +51.590\left(\alp r_0^2\right) +2753\left(\alp r_0^2\right)^2} \,,\label{eq:Fit_alp_emDir_1001}
\ea
\ba
^{\bfs{EM}}F^{10}_{0.001}\left(\alp r_0^2\right)&=& -0.223144\left(\alp r_0^2\right) -0.136142\left(\alp r_0^2\right)^2 \,,\label{eq:Fit_alp_EM_10}\\
^{\bfs{Dir}}F^{10}_{0.001}\left(\alp r_0^2\right)&=& 0.020072\left(\alp r_0^2\right) -0.311515\left(\alp r_0^2\right)^2 \,,\label{eq:Fit_alp_Dir_10}\\
^{\bfs{EM+Dir}}F^{10}_{0.001}\left(\alp r_0^2\right)&=& 0.009448\left(\alp r_0^2\right) -0.303855\left(\alp r_0^2\right)^2 \,.\label{eq:Fit_alp_emDir_10}
\ea

From the above fitting functions (which provide good estimates for the relative difference even for deformations with amplitude of the order of $10\%$ of the main peak), we see that for far-zone deformations, the linear dependence is more prominent compared to near-horizon deformations. Nevertheless in all cases the linear relation in the general laws is evident for amplitudes of $|\alpha\,r_0^2|\lessapprox0.00025$ corresponding to about $1\%$ of the main potential peaks.

From Fig.~\ref{fig:TotalEER_RD_alp} and the above fitting relations, we observe that bumps of similar height, when centered near the event horizon and in the far zone respectively, lead to shifts in the emission rates that differ by three or more orders of magnitude. This indicates that near-horizon deformations are significantly more influential than similar features located in the far zone.

\subsection{Variations of the location of the deformation}
\label{Sec:Variations_of_location}

In order to study the effect that the location of the deformations~\eqref{eq:Gaussian_bump_eps} has on the emission, we fix the height and width parameters of the bump and consider various values for $r_m/r_0$. For the near-horizon deformations, we consider values for $r_m/r_0$ ranging from very near the black-hole horizon up to $r_m=1.1\,r_0$. This restricted range of values ensures that the main peak of the dominant-mode effective potentials (located at $r_p\approx1.21\,r_0$ and $r_p=1.5\,r_0$ for the Dirac and electromagnetic test fields respectively) remains essentially unaffected by the presence of the deformation. To further minimize the impact of the deformation on the main peak, we consider highly-localized near-horizon deformations corresponding to very small values for $\kap\,r_0^{-2}$, much smaller in comparison with the width for the far-zone deformations.

\begin{figure*}[h!]
\begin{center}
\includegraphics[width=0.49\linewidth]{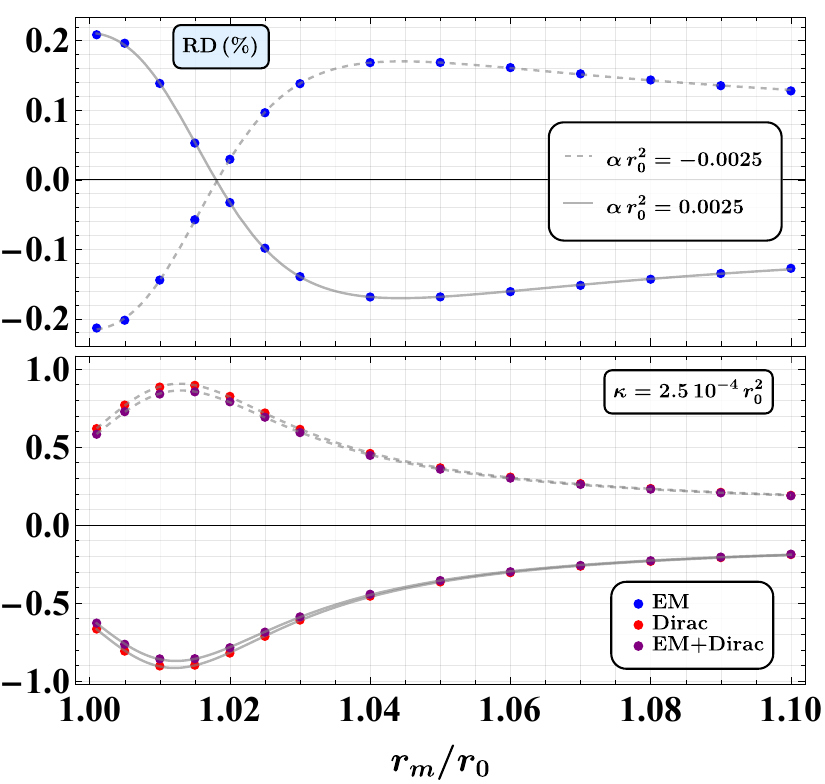}
\includegraphics[width=0.50\linewidth]{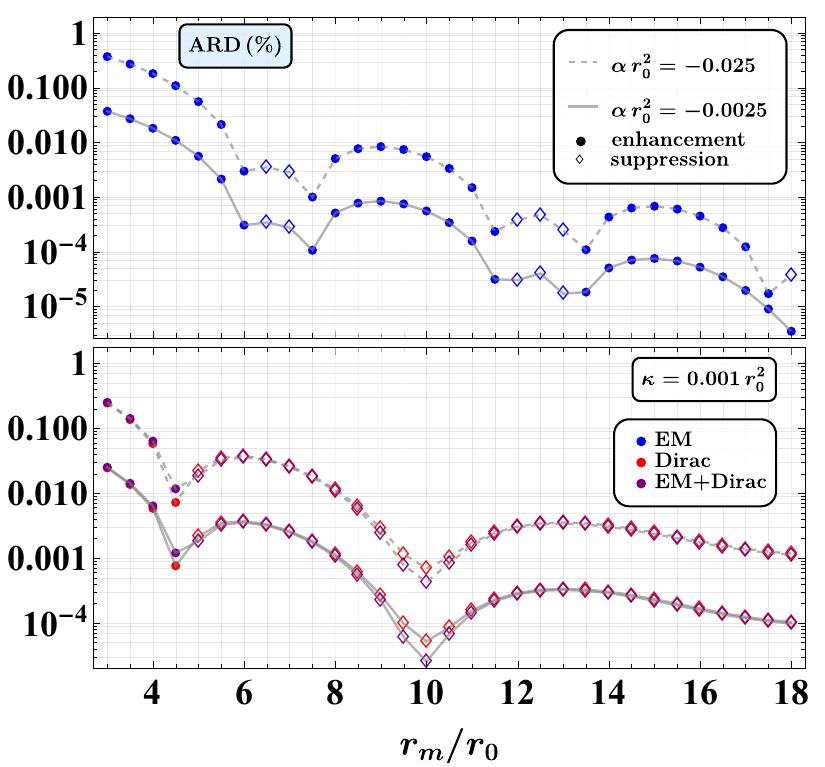}
\includegraphics[width=0.47\linewidth]{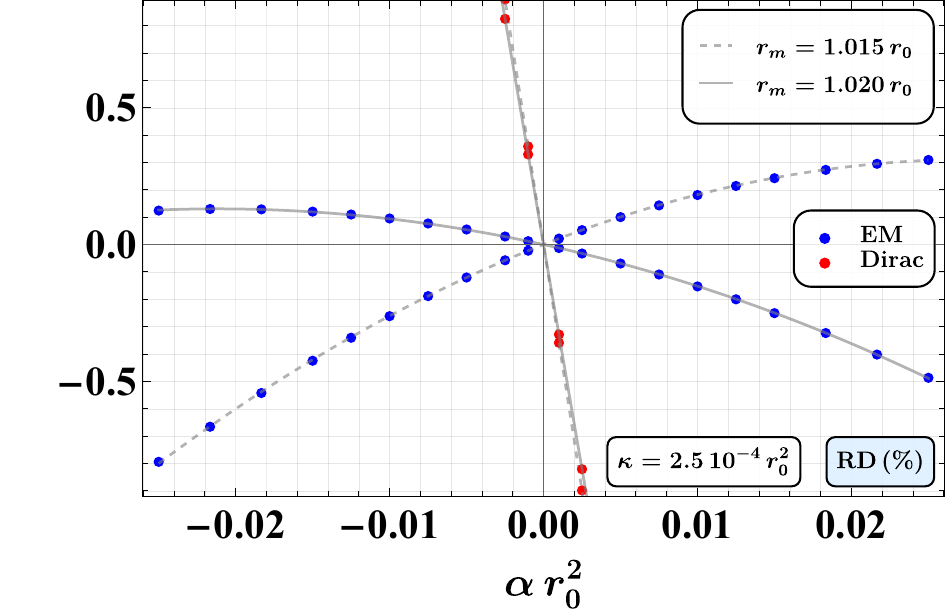}
\includegraphics[width=0.52\linewidth]{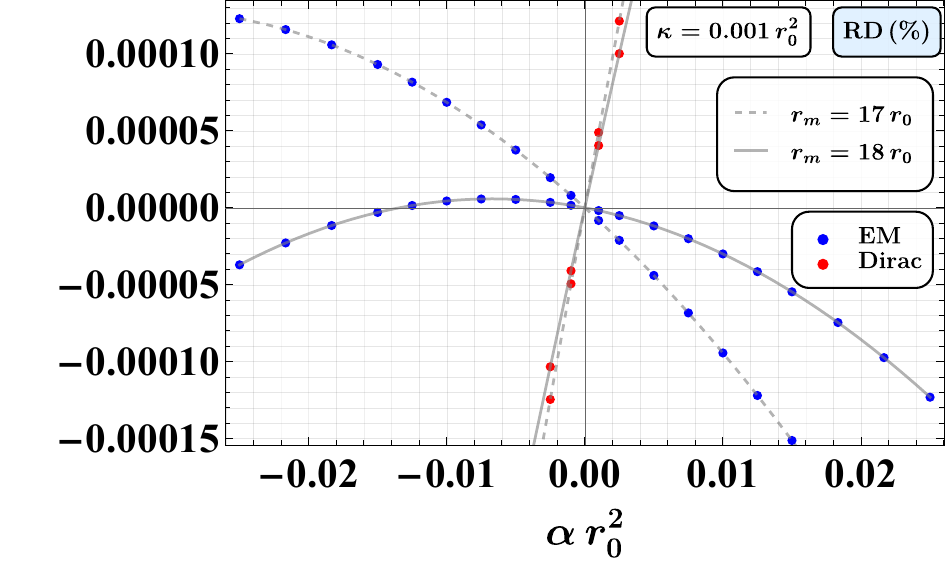}
\caption{Relative,~and absolute relative difference between the total emissivity $r_0^2 \dd E/ \dd t$ in the case of localized deformations~\eqref{eq:Gaussian_bump_eps} and its Schwarzschild limit, demonstrating the impact of the location of the deformations on the emission. Left and right panels correspond to near-horizon and far-zone deformations respectively. The gray-solid and dashed curves correspond to the fitting functions~\eqref{eq:Fit_rm_em_m00025}-\eqref{eq:Fit_rm_emDir_00025}, and~\eqref{eq:Fit_alp_EM_1015}~\eqref{eq:Fit_alp_emDir_18}.~In the top-right panel, the gray-solid and dashed curves are used only to distinguish between data sets and do not correspond to fits of the data.}
\label{fig:TotalEER_ARD_rm}
\end{center}
\end{figure*}

According to the data of table~\ref{tab:2}, see also top-left panel of Fig.~\ref{fig:TotalEER_ARD_rm}, for near-horizon deformations with a height parameter $\alp\,r_0^2$ corresponding to about $1\%$ of the main peak of the effective potentials, the location of the bump has significant implications for the emission that depend on the type of the emitted particles.

\begin{table}
\centering
\resizebox{\textwidth}{!}{
\begin{tabular}{|c||c|c||c|c||c|c||}
\cline{2-7}
\multicolumn{1}{c||}{} & \multicolumn{2}{c||}{EM} & \multicolumn{2}{c||}{Dirac} & \multicolumn{2}{c||}{EM $+$ Dirac} \\
\hline\hline
\multicolumn{1}{|c||}{$r_m/r_0$} & \multicolumn{1}{c|}{$\alp\,r_0^{2}=-0.0025$} & \multicolumn{1}{c||}{$\alp\,r_0^{2}=0.0025$} & \multicolumn{1}{c|}{$\alp\,r_0^{2}=-0.0025$} & \multicolumn{1}{c||}{$\alp\,r_0^{2}=0.0025$} & \multicolumn{1}{c|}{$\alp\,r_0^{2}=-0.0025$} & \multicolumn{1}{c||}{$\alp\,r_0^{2}=0.0025$} \\
\hline
\hline
1.001 & -0.213456~$\%$ & 0.208471~$\%$ & 0.618183~$\%$ & -0.666984~$\%$ & 0.581858~$\%$ & -0.628745~$\%$ \\
1.01  & -0.144506~$\%$ & 0.138455~$\%$ & 0.884525~$\%$ & -0.903712~$\%$ & 0.839578~$\%$ & -0.858192~$\%$ \\
1.02  & 0.0293169~$\%$ & -0.0329382~$\%$ & 0.824721~$\%$ & -0.821136~$\%$ & 0.789979~$\%$ & -0.786708~$\%$ \\
1.03  & 0.138062~$\%$ & -0.139466~$\%$ & 0.613351~$\%$ & -0.609441~$\%$ & 0.592591~$\%$ & -0.588913~$\%$ \\
1.05  & 0.168613~$\%$ & -0.168689~$\%$ & 0.366582~$\%$ & -0.364982~$\%$ & 0.357935~$\%$ & -0.356408~$\%$ \\
1.07  & 0.152249~$\%$ & -0.152128~$\%$ & 0.265045~$\%$ & -0.264175~$\%$ & 0.260119~$\%$ & -0.259281~$\%$ \\
1.1   & 0.127692~$\%$ & -0.127547~$\%$ & 0.190019~$\%$ & -0.18956~$\%$  & 0.187297~$\%$ & -0.186851~$\%$ \\
\hline\hline
\end{tabular}
}
\caption{Relative difference between the total emissivity in the presence of various near-horizon localized deformations~\eqref{eq:Gaussian_bump_eps} of the effective potential with $\kap=2.5\times10^{-4}\,r_0^2$, and the undeformed limits of $r_0^2\,dE/dt\simeq 0.000134554185$, and $r_0^2\,\dd E/\dd t \simeq 0.002945989552$ for the EM and Dirac channels respectively. Positive (negative) values indicate enhancement (suppression) of the emission due to the deformation. The contributions from the the first four multipole numbers have been taken into account. See also, Fig.~\ref{fig:TotalEER_ARD_rm}.}
\label{tab:2}
\end{table}

In particular we observe that for the fermionic sector a positive bump suppresses the emission and a positive bump enhances it. For bumps that are located very near the event horizon, the bosonic sector exhibits the exact opposite response with e.g. positive bumps enhancing the emission. However, as the location of the bump is taken gradually at larger radii, the aforementioned enhancement becomes milder and ceases at a critical radius of $r_m\simeq 1.018\,r_0$. The value for the critical radius can been inferred from the fitting functions~\eqref{eq:Fit_rm_em_m00025} and~\eqref{eq:Fit_rm_em_00025}.  For the same positive bump located at even larger radii, the emission of bosons becomes suppressed with respect to the Schwarzschild value. This inversion of the bosonic sector's response to the deformations holds for any value of the amplitude parameter as can be seen in the bottom-left panel of Fig.~\ref{fig:TotalEER_ARD_rm}.

\begin{figure*}[h!]
\begin{center}
\includegraphics[width=0.49\linewidth]{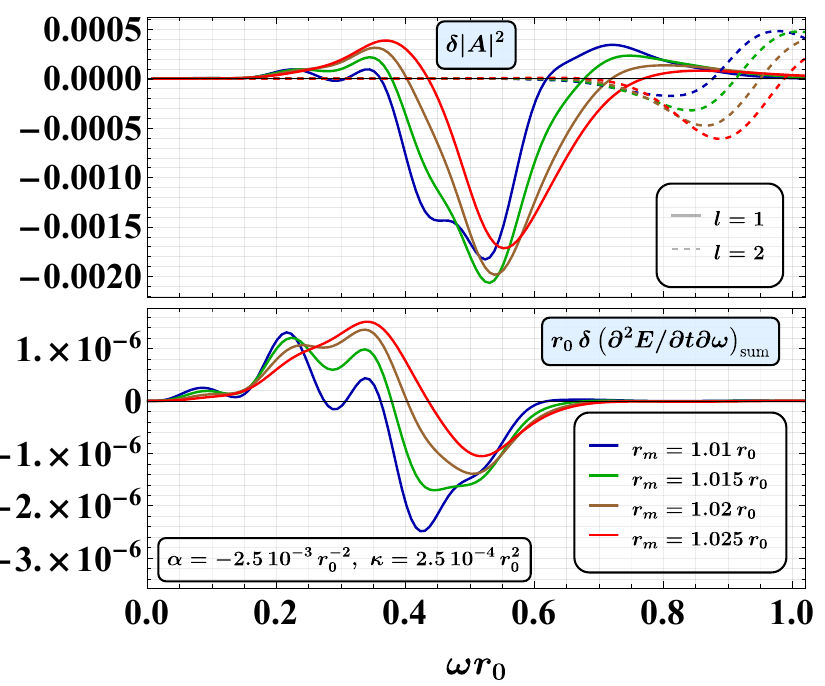}
\includegraphics[width=0.50\linewidth]{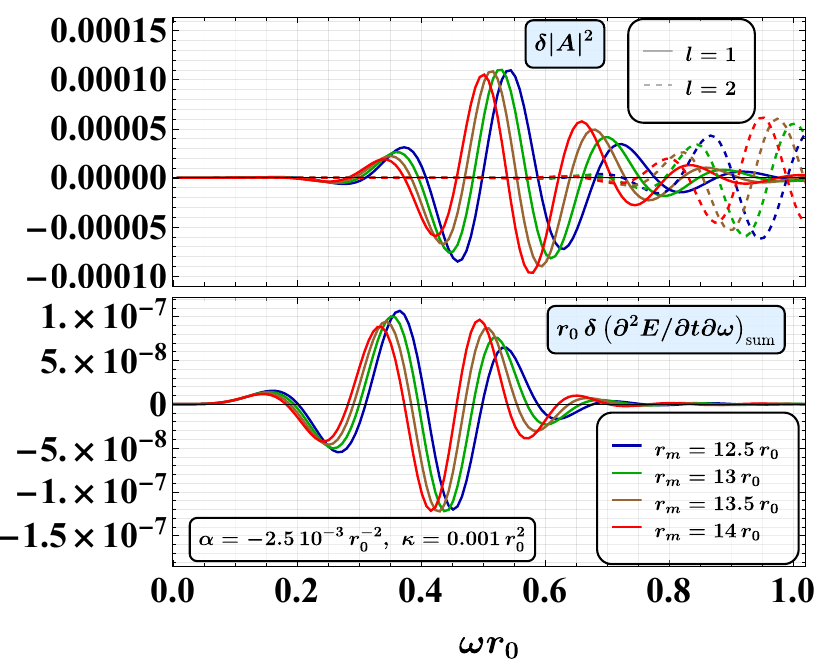}
\caption{The effect of the location of the deformation~\eqref{eq:Gaussian_bump_eps} for the bosonic channel of emission.~For details on the contents of the panels, see caption of Fig.~\ref{fig:DeltaGFandEER_alpha_variations}.}
\label{fig:DeltaGFandEER_rm_variations}
\end{center}
\end{figure*}

The vanishing of the difference in the total emissivity from the Schwarzschild value for deformations located at some critical radius, can be understood in terms of the differences of the dominant-mode (lowest multipole) grey-body factors and the corresponding EERs as shown in the left panels of Fig.~\ref{fig:DeltaGFandEER_rm_variations}. As the value of $r_m$ becomes larger, there is a gradual increase in the enhancement of emission for low-frequency bosons up to a critical point where intermediate frequency suppression is exactly balanced resulting in no difference for the total emission from the Schwarzschild limit.

For both channels of emission, as the location of the near-horizon deformation moves towards the main peak of the potential, the deviations from the Schwarzschild limit become milder. On the other hand, for far-zone deformations we observe the opposite behavior with the strongest deviations from Schwarzschild corresponding to bumps that are located near the main peak, see table~\ref{tab:3}, and right panels of Fig.~\ref{fig:TotalEER_ARD_rm}. Thus, we conclude that the near-horizon geometry is a more important factor than the mere distance between the main peak and the deformation.

\begin{table}
\centering
\resizebox{\textwidth}{!}{
\begin{tabular}{|c||c|c||c|c||c|c||}
\cline{2-7}
\multicolumn{1}{c||}{} & \multicolumn{2}{c||}{EM} & \multicolumn{2}{c||}{Dirac} & \multicolumn{2}{c||}{EM $+$ Dirac} \\
\hline\hline
\multicolumn{1}{|c||}{$r_m/r_0$} & \multicolumn{1}{c|}{$\alp\,r_0^{2}=-0.025$} & \multicolumn{1}{c||}{$\alp\,r_0^{2}=-0.0025$} & \multicolumn{1}{c|}{$\alp\,r_0^{2}=-0.025$} & \multicolumn{1}{c||}{$\alp\,r_0^{2}=-0.0025$} & \multicolumn{1}{c|}{$\alp\,r_0^{2}=-0.025$} & \multicolumn{1}{c||}{$\alp\,r_0^{2}=-0.0025$} \\
\hline
\hline
3 & 0.372075~$\%$ & 0.037131~$\%$ & 0.245867~$\%$ & 0.024584~$\%$ & 0.25138~$\%$ & 0.0251323~$\%$ \\
4 & 0.182098~$\%$ & 0.018198~$\%$ & 0.058159~$\%$ & 0.005853~$\%$ & 0.0635721~$\%$ & 0.00639226~$\%$ \\
5 & 0.055694~$\%$ & 0.005577~$\%$ & -0.021974~$\%$ & -0.002161~$\%$ & -0.0185817~$\%$ & -0.00182285~$\%$ \\
6 & 0.002978~$\%$ & 0.000307~$\%$ & -0.037468~$\%$ & -0.003718~$\%$ & -0.0357012~$\%$ & -0.0035424~$\%$ \\
7 & -0.002828~$\%$ & -0.000276~$\%$ & -0.026257~$\%$ & -0.002603~$\%$ & -0.0252336~$\%$ & -0.00250169~$\%$ \\
8 & 0.005054~$\%$ & 0.000512~$\%$ & -0.011554~$\%$ & -0.001136~$\%$ & -0.0108283~$\%$ & -0.00106432~$\%$ \\
9 & 0.008371~$\%$ & 0.000844~$\%$ & -0.002881~$\%$ & -0.000270~$\%$ & -0.00238918~$\%$ & -0.000221522~$\%$ \\
10 & 0.005494~$\%$ & 0.000557~$\%$ & -0.000697~$\%$ & -0.000052~$\%$ & -0.000426229~$\%$ & -0.0000255002~$\%$ \\
11 & 0.001488~$\%$ & 0.000157~$\%$ & -0.001730~$\%$ & -0.000156~$\%$ & -0.00158912~$\%$ & -0.00014203~$\%$ \\
12 & -0.000380~$\%$ & -0.000030~$\%$ & -0.003068~$\%$ & -0.000290~$\%$ & -0.00295036~$\%$ & -0.000278598~$\%$ \\
13 & -0.000245~$\%$ & -0.000017~$\%$ & -0.003540~$\%$ & -0.000338~$\%$ & -0.00339564~$\%$ & -0.000323685~$\%$ \\
14 & 0.000431~$\%$ & 0.000050~$\%$ & -0.003195~$\%$ & -0.000304~$\%$ & -0.0030369~$\%$ & -0.00028835~$\%$ \\
15 & 0.000680~$\%$ & 0.000075~$\%$ & -0.002491~$\%$ & -0.000234~$\%$ & -0.00235224~$\%$ & -0.000220325~$\%$ \\
16 & 0.000450~$\%$ & 0.000052~$\%$ & -0.001827~$\%$ & -0.000168~$\%$ & -0.00172738~$\%$ & -0.00015817~$\%$ \\
17 & 0.000123~$\%$ & 0.000020~$\%$ & -0.001392~$\%$ & -0.000125~$\%$ & -0.001326~$\%$ & -0.000118281~$\%$ \\
18 & -0.000037~$\%$ & 0.000003~$\%$ & -0.001178~$\%$ & -0.000103~$\%$ & -0.00112789~$\%$ & -0.0000986715~$\%$ \\
\hline\hline
\end{tabular}
}
\caption{Relative difference between the total emissivity in the presence of various far-zone localized deformations~\eqref{eq:Gaussian_bump_eps} of the effective potential with $\kap=0.001\,r_0^2$, and the undeformed limits of $r_0^2\,dE/dt\simeq 0.000134554185$, and $r_0^2\,\dd E/\dd t \simeq 0.002945989552$ for the EM and Dirac channels respectively. Positive (negative) values indicate enhancement (suppression) of the emission due to the deformation.~The contributions from the the first four multipole numbers have been taken into account. See also, Fig.~\ref{fig:TotalEER_ARD_rm}.}
\label{tab:3}
\end{table}

In the case of far-zone deformations with a negative amplitude, we see that both channels respond with enhancement of the total emission when the bump is located near ISCO. This can be attributed to the lowering of part of the tail of the potential barrier due to the dips. However, as $r_m$ increases, the fermionic sector transitions to suppression of the emission for any value of $r_m$ while the emission of bosons exhibits enhancement, suppression or no deviation from the Schwarzschild limit depending in the value of $r_m$, as can be seen in top-right panel of Fig.~\ref{fig:TotalEER_ARD_rm}. This phenomenon can be once again understood in terms of the balance between enhancement and suppression of the energy emission rate at various frequencies as shown in the right panel of Fig.~\ref{fig:DeltaGFandEER_rm_variations}.

\begin{figure*}[h!]
\centering
\begin{minipage}{0.53\textwidth}
\centering
\includegraphics[width=\linewidth]{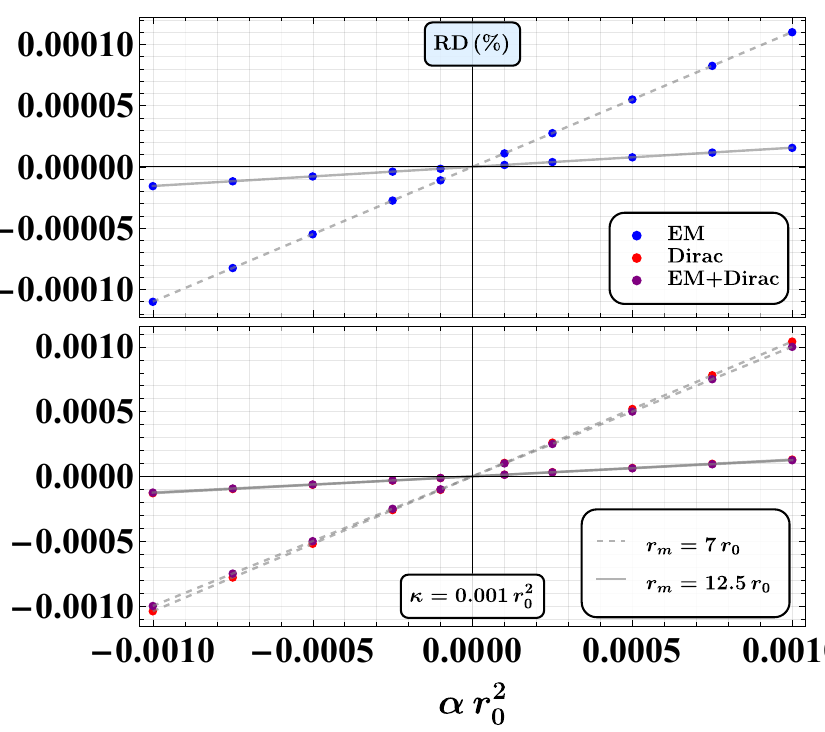}
\end{minipage}
\hfill
\begin{minipage}{0.45\textwidth}
\centering
\textbf{Slopes for the fitting lines} \\[0.5em]
\begin{tabular}{||c||c|c||}
\hline
& $r_m=7\,r_0$ & $r_m=12.5\,r_0$ \\
\hline\hline
EM & 0.110051 & 0.015632 \\
\hline
Dirac & 1.040347 & 0.129702 \\
\hline
EM $+$ Dirac & 0.999713 & 0.124720 \\
\hline\hline
\end{tabular}
\end{minipage}
\caption{Relative difference between the total emissivity $r_0^2 \dd E/ \dd t$ and its Schwarzschild limit, for localized deformations~\eqref{eq:Gaussian_bump_eps} with $\alp\,r_0^2\ll1$ in the far zone. The gray lines correspond to fits of the numerical data for which the slopes are given in the adjacent table.}
\label{fig:TotalEER_RD_alp_[rm7_125]}
\end{figure*}

The observed transition from enhancement to suppression for the emission of bosons can be of two types, depending on the parameters of the deformation. First, we have the transitions that depend solely to the value of $r_m$, and are independent of the amplitude of the deformation. Such examples are given in the neighborhoods of $r_m/r_0=7,12$ in the top-right panel of Fig.~\ref{fig:TotalEER_ARD_rm}, see also~Fig.~\ref{fig:TotalEER_RD_alp_[rm7_125]}. A second type of transition can be observed at $r_m/r_0=18$ where the amplitude of the deformation changes the response of the emission, as can be seen in the bottom-right panel in~Fig.~\ref{fig:TotalEER_ARD_rm}.

In equations~\eqref{eq:Fit_rm_em_m00025}-\eqref{eq:Fit_rm_emDir_00025}, we provide the fitting functions for the numerical data presented in the top-left panel of Fig.~\ref{fig:TotalEER_ARD_rm} for the near-horizon deformations. The right superscript in $F$, denotes the value of $\alp\,r_0^2$, while the subscript indicates the value of $\kap\,r_0^{-2}$. The values of $F(r_m/r_0)$ correspond to the percentage of relative difference of the total emissivity for dimensionless parameter $r_m/r_0$ and the Schwarzschild limit.
\ba
^{\bfs{EM}}F^{-0.0025}_{0.00025}\left(\frac{r_m}{r_0}\right)&=& \frac{0.0550906 -0.1157353\left(r_m/r_0\right) +0.0605275\left(r_m/r_0\right)^2}{1 -1.9749360\left(r_m/r_0\right) +0.9754781\left(r_m/r_0\right)^2} \,,\label{eq:Fit_rm_em_m00025}\\
^{\bfs{EM}}F^{0.0025}_{0.00025}\left(\frac{r_m}{r_0}\right)&=& \frac{-0.0549200 +0.1153851\left(r_m/r_0\right) -0.0603499\left(r_m/r_0\right)^2}{1 -1.9749859\left(r_m/r_0\right) +0.9755312\left(r_m/r_0\right)^2} \,,\label{eq:Fit_rm_em_00025}\\
^{\bfs{Dir}}F^{-0.0025}_{0.00025}\left(\frac{r_m}{r_0}\right)&=& \frac{0.0727302 -0.1514153\left(r_m/r_0\right) +0.0789260\left(r_m/r_0\right)^2}{1 -1.983198\left(r_m/r_0\right) +0.9836063\left(r_m/r_0\right)^2} \,,\label{eq:Fit_rm_Dir_m00025}
\ea
\ba
^{\bfs{Dir}}F^{0.0025}_{0.00025}\left(\frac{r_m}{r_0}\right)&=& \frac{-0.0733575 +0.1525070\left(r_m/r_0\right) -0.0794169\left(r_m/r_0\right)^2}{1 -1.9843068\left(r_m/r_0\right) +0.9847303\left(r_m/r_0\right)^2} \,,\label{eq:Fit_rm_Dir_00025}\\
^{\bfs{EM+Dir}}F^{-0.0025}_{0.00025}\left(\frac{r_m}{r_0}\right)&=& \frac{0.0716759 -0.1493562\left(r_m/r_0\right) +0.0779078\left(r_m/r_0\right)^2}{1 -1.9833162\left(r_m/r_0\right) +0.9837324\left(r_m/r_0\right)^2} \,,\label{eq:Fit_rm_emDir_m00025}\\
^{\bfs{EM+Dir}}F^{0.0025}_{0.00025}\left(\frac{r_m}{r_0}\right)&=& \frac{-0.0721509 +0.1501654\left(r_m/r_0\right) -0.0782676\left(r_m/r_0\right)^2}{1 -1.9844404\left(r_m/r_0\right) +0.9848662\left(r_m/r_0\right)^2} \,.\label{eq:Fit_rm_emDir_00025}
\ea

In addition to providing a very good estimate for the relative difference in total emissivity at any value of $r_m/r_0$ in the near-horizon regime, and allowing us to determine the critical radius corresponding to the absence of deviation from the Schwarzschild limit, the numerical values of the coefficients in the above fitting functions provide further demonstration of the linear response of the emission for relatively small amplitudes of the deformations. Even though we have not been able to obtain equivalent compact analytic expressions for the fit of the numerical data for the far-zone deformations, the main observation in that case is that the difference in total emissivity from the Schwarzschild limit becomes smaller the further away the deformation is located from the peak of the effective potential.

\subsection{Variations of the width of the deformation}
\label{Sec:Variations_of_width}

In this subsection, we study the effect of the width of the deformations on the emission by fixing the values for the height and location parameters in~\eqref{eq:Gaussian_bump_eps}, and considering different values for $\kap\,r_0^{-2}$. In order to ensure, once again, that the impact of the near-horizon deformations on the main peak of the effective potential is not significant, we restricted the maximum value of the width parameter to $\kap\,r_0^{-2}=0.01$ for the considered amplitude $|\alp\,r_0^2|=0.0025$ (corresponding to about $1\%$ of the main peak of the effective potentials) of the deformations. Larger amplitudes would require a lower upped bound on the range of values for the width parameter, see Fig.~\ref{fig:Gaussian_bumps}. In our analysis we have used the same amplitudes and range of values for the width parameter in both near-horizon and far-zone deformations in order to facilitate the comparison between the two cases.

According to the data presented in table~\ref{tab:4}, when the deformations are sharply localized (what corresponds to small values of $\kap\,r_0^{-2}$) near the event horizon, the sign of the amplitude of the deformation affects the two channels of emission in the opposite way, e.g.~a positive bump enhances the emission of bosons and suppresses the emission of fermions. This observation is consistent with the findings presented in the previous section.

\begin{table}
\centering
\resizebox{\textwidth}{!}{
\begin{tabular}{|c|c||c|c||c|c||c|c||}
\cline{3-8}
\multicolumn{2}{c||}{} & \multicolumn{2}{c||}{EM} & \multicolumn{2}{c||}{Dirac} & \multicolumn{2}{c||}{EM $+$ Dirac} \\
\hline\hline
& $\kap\,r_0^{-2}$ & $r_m=1.001\,r_0$ & $r_m=10\,r_0$ & $r_m=1.001\,r_0$ & $r_m=10\,r_0$ & $r_m=1.001\,r_0$ & $r_m=10\,r_0$ \\
\hline
\multirow{10}{*}{\rotatebox[origin=c]{90}{$\alp\,r_0^{2}=0.0025$}} 
& 0.00025 & 0.208471~$\%$ & -0.000279135~$\%$ & -0.666984~$\%$ & 0.0000245628~$\%$ & -0.628745~$\%$ & 0.0000112976~$\%$ \\
& 0.0005 & 0.186029~$\%$ & -0.000394899~$\%$ & -0.828488~$\%$ & 0.0000344658~$\%$ & -0.784175~$\%$ & 0.0000157117~$\%$ \\
& 0.00075 & 0.163589~$\%$ & -0.000483763~$\%$ & -0.929869~$\%$ & 0.0000419611~$\%$ & -0.882108~$\%$ & 0.0000189981~$\%$ \\
& 0.001 & 0.143102~$\%$ & -0.00055871~$\%$ & -1.00464~$\%$ & 0.0000482116~$\%$ & -0.954511~$\%$ & 0.000021702~$\%$ \\
& 0.0025 & 0.050814~$\%$ & -0.000884126~$\%$ & -1.25695~$\%$ & 0.0000746306~$\%$ & -1.19983~$\%$ & 0.0000327534~$\%$ \\
& 0.005 & -0.0474753~$\%$ & -0.00125146~$\%$ & -1.46049~$\%$ & 0.000103146~$\%$ & -1.39877~$\%$ & 0.0000439781~$\%$ \\
& 0.0075 & -0.116455~$\%$ & -0.00153373~$\%$ & -1.5841~$\%$ & 0.000124207~$\%$ & -1.52~$\%$ & 0.0000517905~$\%$ \\
& 0.01 & -0.170486~$\%$ & -0.00177194~$\%$ & -1.67378~$\%$ & 0.000141464~$\%$ & -1.60812~$\%$ & 0.0000578884~$\%$ \\
\hline
\multirow{10}{*}{\rotatebox[origin=c]{90}{$\alp\,r_0^{2}=-0.0025$}} 
& 0.00025 & -0.213456~$\%$ & 0.00027871~$\%$ & 0.618183~$\%$ & -0.0000255365~$\%$ & 0.581858~$\%$ & -0.0000122474~$\%$ \\
& 0.0005 & -0.192292~$\%$ & 0.000394048~$\%$ & 0.781071~$\%$ & -0.0000364131~$\%$ & 0.738555~$\%$ & -0.0000176111~$\%$ \\
& 0.00075 & -0.170695~$\%$ & 0.000482487~$\%$ & 0.883767~$\%$ & -0.0000448818~$\%$ & 0.837709~$\%$ & -0.000021847~$\%$ \\
& 0.001 & -0.150837~$\%$ & 0.000557008~$\%$ & 0.959715~$\%$ & -0.0000521055~$\%$ & 0.911208~$\%$ & -0.0000255002~$\%$ \\
& 0.0025 & -0.0606273~$\%$ & 0.000879873~$\%$ & 1.2172~$\%$ & -0.0000843597~$\%$ & 1.16139~$\%$ & -0.0000422433~$\%$ \\
& 0.005 & 0.0361943~$\%$ & 0.00124296~$\%$ & 1.42622~$\%$ & -0.000122585~$\%$ & 1.3655~$\%$ & -0.0000629397~$\%$ \\
& 0.0075 & 0.104448~$\%$ & 0.001521~$\%$ & 1.55373~$\%$ & -0.000153339~$\%$ & 1.49042~$\%$ & -0.0000802057~$\%$ \\
& 0.01 & 0.158057~$\%$ & 0.001755~$\%$ & 1.64649~$\%$ & -0.000180269~$\%$ & 1.58148~$\%$ & -0.0000957393~$\%$ \\
\hline\hline
\end{tabular}
}
\caption{Relative difference between the total emissivity in the presence of various types of Gaussian-bump deformations of the effective potential~\eqref{eq:Gaussian_bump}, and the undeformed limit. Positive (negative) values indicate enhancement (suppression) of the emission due to the deformation. The contributions from the the first four multipole numbers have been taken into account. See also, Fig.~\ref{fig:TotalEER_RD_kap}.}
\label{tab:4}
\end{table}

\begin{figure*}[h!]
\begin{center}
\includegraphics[width=0.47\linewidth]{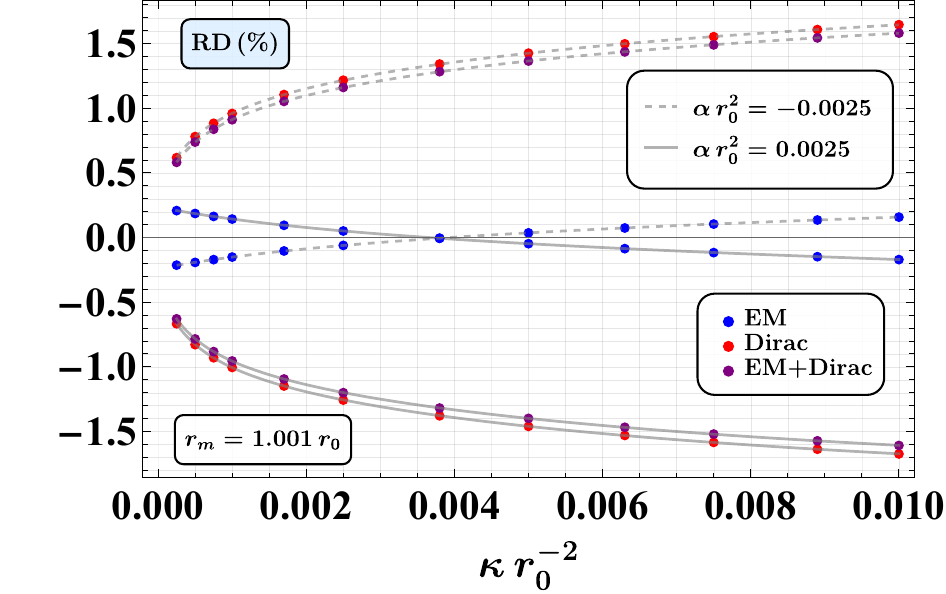}
\includegraphics[width=0.51\linewidth]{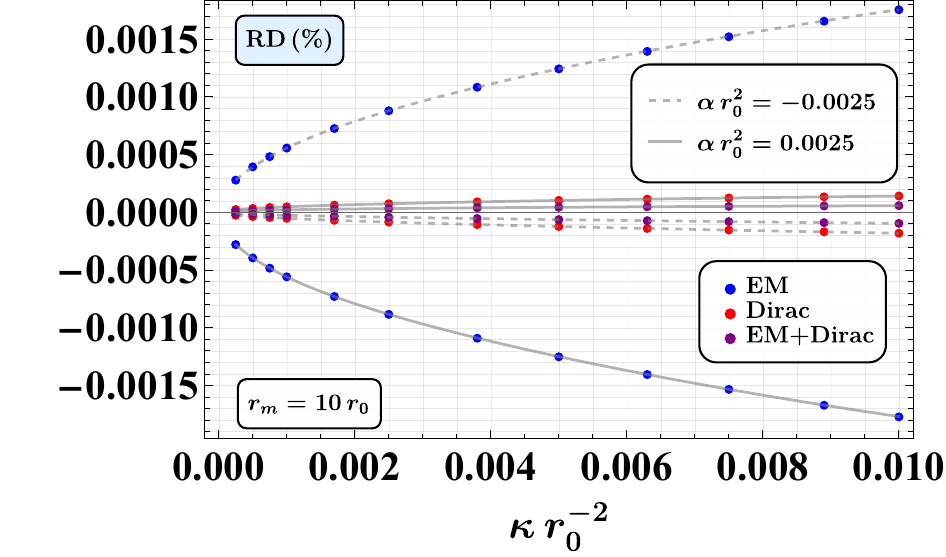}
\caption{Relative difference between the total emissivity $r_0^2 \dd E/ \dd t$ in the case of localized deformations~\eqref{eq:Gaussian_bump_eps} and its Schwarzschild limit, demonstrating the impact of the width of the deformations on the emission.~The gray-solid and dashed curves correspond to the fitting functions~\eqref{eq:Fit_kap_em_1001m}-\eqref{eq:Fit_rm_emDir_10}.}
\label{fig:TotalEER_RD_kap}
\end{center}
\end{figure*}

As is shown in the left panel of Fig.~\ref{fig:TotalEER_RD_kap}, in the case of the fermionic sector, increasing the width of the near-horizon deformations results in a further increase in the difference of the total energy emission rate from the Schwarzschild limit while preserving the type of modification i.e. enhancement for negative bumps and suppression for positive bumps.

\begin{figure*}[h!]
\begin{center}
\includegraphics[width=0.49\linewidth]{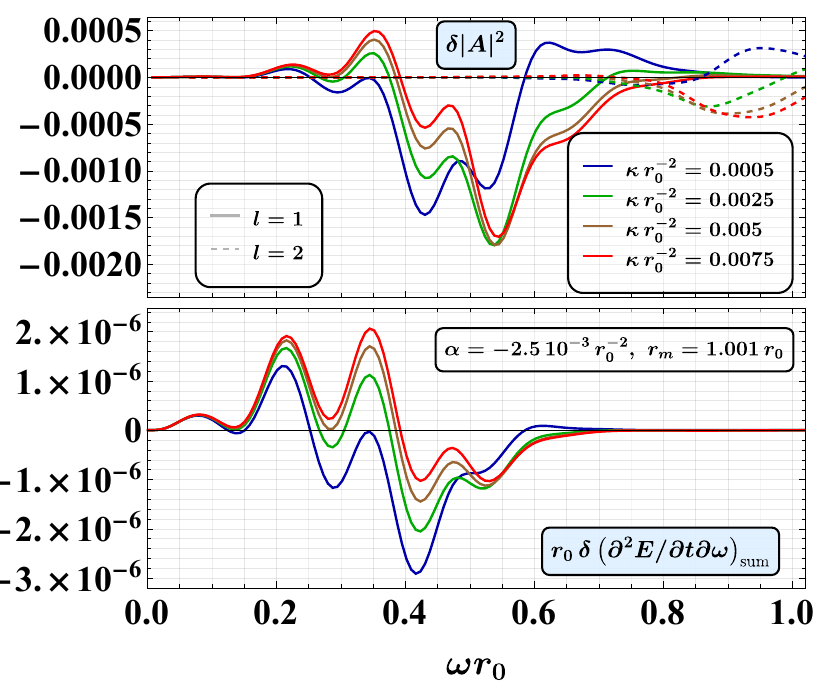}
\includegraphics[width=0.49\linewidth]{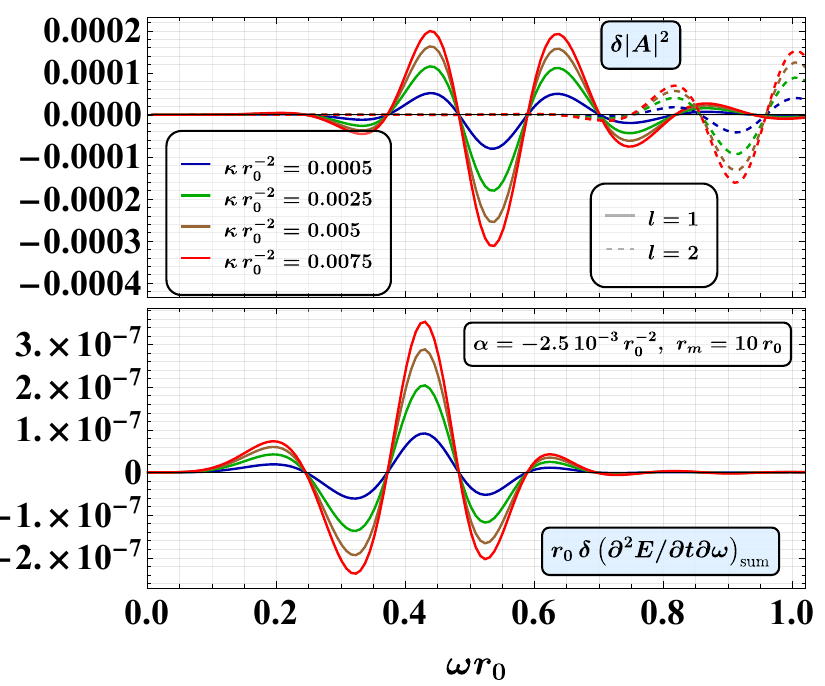}
\caption{The effect of the width of the deformation~\eqref{eq:Gaussian_bump_eps} for the bosonic channel of emission.~For details on the contents of the panels, see caption of Fig.~\ref{fig:DeltaGFandEER_alpha_variations}.}
\label{fig:DeltaGFandEER_kappa_variations}
\end{center}
\end{figure*}

The response of the bosonic sector is again distinct from, and more intricate than, the fermionic one, with the difference from the corresponding Schwarzschild value initially decreasing with the width of the bump, until a critical width where it vanishes, and subsequently it increases for broader deformations albeit with the opposite response i.e. enhancement turns to suppression and vise versa. This is again a consequence of the interplay between enhancement and suppression of the power spectrum at various frequencies as shown on the left panel of Fig.~\ref{fig:DeltaGFandEER_kappa_variations}.

The critical width for the parameter values of our example, can be inferred from the fitting functions~\eqref{eq:Fit_kap_em_1001m} and~\eqref{eq:Fit_kap_em_1001} and corresponds to $\kap\,r_0^{-2}\simeq0.0037$. As such, we find that when the near horizon deformations are sufficiently broad (without affecting significantly the main peak), both channels of emission exhibit the same type of response to the deformations.

In the right panel of Fig.~\ref{fig:TotalEER_RD_kap} we observe that for far-zone deformations (and for the indicative value of $r_m=10\,r_0$), the differences from the Schwarzschild limit grow monotonically with the width of the bumps for both types of particles. Furthermore, the two channels respond oppositely to the amplitude of the deformation throughout the range of considered values for the width parameter. We also observe that when the width of the deformations is sufficiently large, the response of both channels approaches an approximately linear dependence on $\kap\,r_0^{-2}$.

The aforementioned dependence of the total emission of the black hole on the width of the deformations can be understood in terms of the impact of the width parameter on the frequency profiles for the GFs (and consequently on the power spectra) as shown in Fig.~\ref{fig:DeltaGFandEER_kappa_variations}. In particular, we observe that for far-zone deformations, the variation of the width parameter preserves the sign for the difference of the GFs from the Schwarzschild limit for all frequencies. This property is similar to the effect of the amplitude of the bump, that also preserves the sign for the difference at any given frequency, see Fig.~\ref{fig:DeltaGFandEER_alpha_variations}. On the other hand, the variation of the location parameter for example, induces also a shift on the frequency profiles of the GF differences, as shown on the right panel of~Fig.~\ref{fig:DeltaGFandEER_rm_variations}, and as such, the sign of the difference is not preserved for all frequencies. This property provides a better understanding for the observed response of the total emission on the various parameters of the deformations as shown in Figs.~\ref{fig:TotalEER_RD_alp},~\ref{fig:TotalEER_ARD_rm}, and~\ref{fig:TotalEER_RD_kap}. When there is no sign change for the differences in the GFs at any given frequency, the type of response of the total emissivity (enhancement or suppression) does not change~\footnote{It should be noted however that the sign-change of the differences of the GFs is not a sufficient condition for the change in the response of the total emission, since this also depends in general on the way that the amplitudes of the differences at any given frequency change with the variation of the bump parameters.}. On the other hand, sign-changes on the difference of GFs may result in a different response for the total emission since the sign of the integral over all frequencies of the power spectrum may change, e.g. compare the left and right panels of Figs.~\ref{fig:DeltaGFandEER_kappa_variations} and~\ref{fig:TotalEER_RD_kap}. 

In equations~\eqref{eq:Fit_kap_em_1001m}-\eqref{eq:Fit_rm_emDir_10}, we provide the fitting functions $F$ for the numerical data of Fig.~\ref{fig:TotalEER_RD_kap}. The right superscript in $F$, denotes the value of $r_m/r_0$, while the subscript indicates the value of $\alp r_0^2$. The values of $F(\kap\,r_0^2)$ correspond to the percentage of relative difference of the total emissivity for dimensionless parameter $\kap\,r_0^2$ and the Schwarzschild limit.

\ba
^{\bfs{EM}}F^{1.001}_{-0.0025}\left(\kap\,r_0^{-2}\right)&=& \frac{-0.2397477 +50.61 \left(\kap\,r_0^{-2}\right) +2641\left(\kap\,r_0^{-2}\right)^2}{1 +235\left(\kap\,r_0^{-2}\right)} \,,\label{eq:Fit_kap_em_1001m}\\
^{\bfs{Dir}}F^{1.001}_{-0.0025}\left(\kap\,r_0^{-2}\right)&=& \frac{0.3155578 +2496\left(\kap\,r_0^{-2}\right) +500953\left(\kap\,r_0^{-2}\right)^2}{1 +2221\left(\kap\,r_0^{-2}\right) +225985\left(\kap\,r_0^{-2}\right)^2} \,,\label{eq:Fit_rm_Dir_1001m}\\
^{\bfs{EM+Dir}}F^{1.001}_{-0.0025}\left(\kap\,r_0^{-2}\right)&=& \frac{0.2928035 +2349\left(\kap\,r_0^{-2}\right) +473315\left(\kap\,r_0^{-2}\right)^2}{1 +2193\left(\kap\,r_0^{-2}\right) +220645\left(\kap\,r_0^{-2}\right)^2} \,,\label{eq:Fit_rm_emDir_1001m}
\ea
\ba
^{\bfs{EM}}F^{1.001}_{0.0025}\left(\kap\,r_0^{-2}\right)&=& \frac{0.2358075 -54\left(\kap\,r_0^{-2}\right) -2876\left(\kap\,r_0^{-2}\right)^2}{1 +248\left(\kap\,r_0^{-2}\right)} \,,\label{eq:Fit_kap_em_1001}\\
^{\bfs{Dir}}F^{1.001}_{0.0025}\left(\kap\,r_0^{-2}\right)&=& \frac{-0.3642567 -2628\left(\kap\,r_0^{-2}\right) -518640\left(\kap\,r_0^{-2}\right)^2}{1 +2257\left(\kap\,r_0^{-2}\right) +233686\left(\kap\,r_0^{-2}\right)^2} \,,\label{eq:Fit_rm_Dir_1001}\\
^{\bfs{EM+Dir}}F^{1.001}_{0.0025}\left(\kap\,r_0^{-2}\right)&=& \frac{-0.3395421 -2474\left(\kap\,r_0^{-2}\right) -489774\left(\kap\,r_0^{-2}\right)^2}{1 +2228\left(\kap\,r_0^{-2}\right) +228002\left(\kap\,r_0^{-2}\right)^2} \,,\label{eq:Fit_rm_emDir_1001}
\ea
\ba
^{\bfs{EM}}F^{10}_{-0.0025}\left(\kap\,r_0^{-2}\right)&=& \frac{0.0001255 +0.8835712\left(\kap\,r_0^{-2}\right) +203\left(\kap\,r_0^{-2}\right)^2}{1 +1128\left(\kap\,r_0^{-2}\right) +43805\left(\kap\,r_0^{-2}\right)^2} \,,\label{eq:Fit_kap_em_10m}\\
^{\bfs{Dir}}F^{10}_{-0.0025}\left(\kap\,r_0^{-2}\right)&=& \frac{-0.0000115 -0.0793882\left(\kap\,r_0^{-2}\right) -18\left(\kap\,r_0^{-2}\right)^2}{1 +1056\left(\kap\,r_0^{-2}\right) +28861\left(\kap\,r_0^{-2}\right)^2} \,,\label{eq:Fit_rm_Dir_10m}\\
^{\bfs{EM+Dir}}F^{10}_{-0.0025}\left(\kap\,r_0^{-2}\right)&=& \frac{-0.0000054 -0.0377886\left(\kap\,r_0^{-2}\right) -8.85\left(\kap\,r_0^{-2}\right)^2}{1 +1018\left(\kap\,r_0^{-2}\right) +20815\left(\kap\,r_0^{-2}\right)^2} \,,\label{eq:Fit_rm_emDir_10m}
\ea
\ba
^{\bfs{EM}}F^{10}_{0.0025}\left(\kap\,r_0^{-2}\right)&=& \frac{-0.0001256 -0.8836157\left(\kap\,r_0^{-2}\right) -202\left(\kap\,r_0^{-2}\right)^2}{1 +1122\left(\kap\,r_0^{-2}\right) +42638\left(\kap\,r_0^{-2}\right)^2} \,,\label{eq:Fit_kap_em_10}\\
^{\bfs{Dir}}F^{10}_{0.0025}\left(\kap\,r_0^{-2}\right)&=& \frac{0.0000111 +0.0789060\left(\kap\,r_0^{-2}\right) +18\left(\kap\,r_0^{-2}\right)^2}{1 +1182\left(\kap\,r_0^{-2}\right) +56059\left(\kap\,r_0^{-2}\right)^2} \,,\label{eq:Fit_rm_Dir_10}\\
^{\bfs{EM+Dir}}F^{10}_{0.0025}\left(\kap\,r_0^{-2}\right)&=& \frac{0.0000051 +0.0375627\left(\kap\,r_0^{-2}\right) +9.26\left(\kap\,r_0^{-2}\right)^2}{1 +1303\left(\kap\,r_0^{-2}\right) +85626\left(\kap\,r_0^{-2}\right)^2} \,.\label{eq:Fit_rm_emDir_10}
\ea

\section{Possible extensions and generalizations}
\label{Sec:Generalization}

While in this work we considered perturbations of test fields in the background of a Schwarzschild black hole with a bump height that does not depend on the multipole number $\ell$, we can argue that the main conclusions should remain valid in more general situations. A natural extension would be to study gravitational perturbations, taking into account the energy–momentum tensor that could represent the matter responsible for producing the bump. In such a case, a self-consistent treatment of the problem requires solving not the vacuum Einstein equations, but the Einstein equations with an energy–momentum tensor describing the environment or additional matter content. Consequently, the background metric would no longer be the exact Schwarzschild solution, but rather a Schwarzschild metric with small deformations of the form
\begin{equation}
g_{tt}^{\text{Schw.}} \;\rightarrow\; g_{tt}^{\text{Schw.}} + h_{1}(r), 
\quad 
g_{rr}^{\text{Schw.}} \;\rightarrow\; g_{rr}^{\text{Schw.}} + h_{2}(r),
\end{equation}
where $h_{1}(r) \ll g_{tt}$ and $h_{2}(r) \ll g_{rr}$.

Such small static deformations induce modifications of the effective potential, whose explicit form depends on the energy content of the configuration, but which generally takes the form of a second-order differential operator with a new effective potential. This potential may be expressed schematically as
\begin{equation}
V^{\text{Schw.}}(r) \;\rightarrow\; V^{\text{Schw.}}(r) + H\big(h_{1}(r),h_{2}(r)\big) + \mathcal{O}(h_{i}^{2}),
\end{equation}
where $H(h_{1}(r),h_{2}(r))$ plays the role of the bump, now potentially dependent on the multipole number $\ell$. This $\ell$-dependence is the main difference from our treatment, where the bump was independent of $\ell$. 

For the analysis of grey-body factors, this distinction is not essential, since the bump parameters (height, width, and location) could already be arbitrary in our setup. The main difference arises in the spectrum of Hawking radiation, as the relative contributions from different multipoles may change. Nevertheless, we do not expect qualitative modifications of the emission rates, because in most cases the lowest multipoles dominate the radiation, while higher multipoles are strongly suppressed and subdominant. Certainly, such subdominant effects could be studied in detail, but they lie outside the scope of this work, which aims to remain independent of the specific matter content producing the bumps.

This question is intrinsically related to another fundamental issue: how does a bump in the mass function of the background metric translate into the effective potential, and, conversely, how can one reconstruct the metric from a given effective potential? For test fields this may appear to be a relatively simple problem, admitting a naive modification of the form
\begin{equation}
1- \frac{2M}{r} \;\rightarrow\; 1- \frac{2 \big(M+ h(r)\big)}{r},
\end{equation}
where the bump is introduced directly into the mass function. However, a self-consistent introduction of such a mass function implies that the coefficient $g_{rr}$ is modified in this manner, whereas $g_{tt}$ must be chosen so that the full Einstein equations with matter are satisfied, as is done, for example, in models of black holes surrounded by galactic halos \cite{Konoplya:2022hbl}. The solution then again depends on the matter content and its equation of state. 

In general, the reconstruction of the metric from its effective potential is a highly nontrivial problem that can probably be addressed only under a number of simplifying assumptions \cite{Volkel:2019ahb,Albuquerque:2023lzw}. It cannot be accomplished within the agnostic framework that we are pursuing here.

The same concerns regarding the feasibility of an agnostic, theory-independent approach also apply to the thermodynamics of black holes with bumps. In this work we constructed the bumps in such a way that the Hawking temperature remains unchanged, allowing us to isolate the extent to which grey-body factors are affected by the bump and, in turn, influence the Hawking radiation. If, instead, one were to choose a bump that modifies the Hawking temperature, the effect of grey-body factors would become subdominant, making it impossible to disentangle their contribution from the overall radiation intensity.

A consistent formulation of the second law of thermodynamics, as well as the analysis of phase transitions and thermodynamic stability, is only possible when the matter content of the configuration is specified and its equation of state is determined. These issues undoubtedly deserve further study within each particular theory of gravity and for fixed matter content.

Nevertheless, all qualitative conclusions obtained from our general agnostic geometric approach should remain valid, provided we do not assume a strong nonlinear backreaction of the environment on the metric. This robustness is ensured by the arbitrariness of the shape, localization, and height of the bumps considered here.

\section{Conclusions}
\label{Sec:Conclusions}

We have conducted a detailed study of how localized deformations of the effective potential around a Schwarzschild black hole affect grey-body factors and the associated Hawking radiation. These deformations—implemented in a model-independent, Gaussian form—serve to simulate either near-horizon deviations due to quantum or exotic effects, or far-zone perturbations modeling environmental structures such as accretion disks.

Our key findings are:

\begin{itemize}

\item{Near-horizon deformations, even when relatively small, can induce sizable changes in the grey-body factors at low frequencies. This leads to significant variations in the total energy emission rate, particularly for the lowest multipole numbers.}

\item{Environmental (far-zone) deformations have a negligible effect on the Hawking spectrum unless their amplitude is several orders of magnitude larger than physically expected. This supports the idea that typical astrophysical environments around black holes, such as thin accretion disks, do not significantly influence black hole evaporation.}

\item{We have mapped how the emission rates change under systematic variation of bump height, width, and location. While general analytic laws are difficult to formulate, our results provide empirical trends that can serve as useful guidelines.}

\end{itemize}

These results highlight the usefulness of grey-body factors as robust spectral observables, and support their application in probing near-horizon modifications through gravitational wave and Hawking radiation channels.

Our work could be extended in the following way. One could investigate the emission of massive fields, for which the interplay between the effective mass and localized deformations may lead to qualitatively different features in the grey-body spectra, such as additional suppression/enhancement in the low-frequency regime. Another natural direction would be to analyze axially symmetric spacetimes, such as Kerr black holes, where the rotation couples to both the angular momentum and the location of the deformation, potentially amplifying or diminishing the asymmetry between near-horizon and far-zone effects. Finally, it would be worthwhile to examine some particular matter distributions, including non-Gaussian profiles, as well as considering the inclusion of non-gravitational interactions between the environment and the emitted radiation.

In this work, we have considered the grey-body factors and emission rates only for matter test fields. This choice is motivated by the fact that graviton emission depends sensitively on the underlying gravitational theory and the specific energy–momentum distribution responsible for generating the bumps and dips. Since our analysis aims to remain as theory-agnostic as possible, we have focused on fields whose dynamics are governed by well-understood covariant equations. Moreover, for four-dimensional black holes, the emission of gravitons is typically strongly suppressed in comparison to that of matter fields~\cite{Page:1974he,Page:1976df,Konoplya:2019hml,Konoplya:2020cbv}. Nonetheless, once a specific gravitational theory and an equation of state for the perturbing matter are specified, analogous calculations could be extended to gravitational perturbations. Given their similar wave-equation structure, we expect that the qualitative patterns observed here for Maxwell and Dirac fields—particularly the dominant role of near-horizon deformations—would also apply to the gravitational sector.

\begin{acknowledgments}
TDP acknowledges the support of the Research Centre for Theoretical Physics and Astrophysics at the Institute of Physics, Silesian University in Opava. He is particularly grateful to Alexander Zhidenko for insightful discussions on numerical methods and to Euripides Pappas for generously providing computational time on his computer, which significantly expedited the collection of numerical data.
\end{acknowledgments}

\appendix
\section{Fitting functions for numerical data}

The fitting functions $F$ for the numerical data presented in the bottom panels of Fig.~\ref{fig:TotalEER_ARD_rm}, are given in~\eqref{eq:Fit_alp_EM_1015}-\eqref{eq:Fit_alp_emDir_18}. The left superscript in $F$, indicates the channel of emission,~$\bfs{EM}$ for bosons,~$\bfs{Dir}$ for fermions, and~$\bfs{EM+Dir}$ for the combined total emission of both channels. The right superscript, denotes the value of $r_m/r_0$, while the subscript indicates the value of $\kap\,r_0^{-2}$. The values of $F(\alp\,r_0^2)$ correspond to the percentage of relative difference of the total emissivity for dimensionless height parameter $\alp\,r_0^2$ and the Schwarzschild limit.

\ba
^{\bfs{EM}}F^{1.015}_{0.00025}\left(\alp r_0^2\right)&=& 22.156218\left(\alp r_0^2\right) -393.781705\left(\alp r_0^2\right)^2 \,,\label{eq:Fit_alp_EM_1015}\\
^{\bfs{Dir}}F^{1.015}_{0.00025}\left(\alp r_0^2\right)&=& \frac{\left(\alp r_0^2\right) \left[-358.821 -5011\left(\alp r_0^2\right) +18836\left(\alp r_0^2\right)^2\right]}{1 +12.839\left(\alp r_0^2\right)} \,,\label{eq:Fit_alp_Dir_1015}\\
^{\bfs{EM+Dir}}F^{1.015}_{0.00025}\left(\alp r_0^2\right)&=& \frac{\left(\alp r_0^2\right) \left[-342.18 -4781\left(\alp r_0^2\right) +17818\left(\alp r_0^2\right)^2\right]}{1 +12.791\left(\alp r_0^2\right)} \,,\label{eq:Fit_alp_emDir_1015}
\ea
\ba
^{\bfs{EM}}F^{1.02}_{0.00025}\left(\alp r_0^2\right)&=& -12.3034\left(\alp r_0^2\right) -290.305\left(\alp r_0^2\right)^2 \,,\label{eq:Fit_alp_EM_1020}\\
^{\bfs{Dir}}F^{1.02}_{0.00025}\left(\alp r_0^2\right)&=& \frac{\left(\alp r_0^2\right) \left[-329.190 -4670\left(\alp r_0^2\right) +7008\left(\alp r_0^2\right)^2\right]}{1 +15.060\left(\alp r_0^2\right)} \,,\label{eq:Fit_alp_Dir_1020}\\
^{\bfs{EM+Dir}}F^{1.02}_{0.00025}\left(\alp r_0^2\right)&=& \frac{\left(\alp r_0^2\right) \left[-315.356 -4469\left(\alp r_0^2\right) +6514\left(\alp r_0^2\right)^2\right]}{1 +15.004\left(\alp r_0^2\right)} \,,\label{eq:Fit_alp_emDir_1020}
\ea
\ba
^{\bfs{EM}}F^{17}_{0.001}\left(\alp r_0^2\right)&=& -0.008144\left(\alp r_0^2\right) -0.129232\left(\alp r_0^2\right)^2 \,,\label{eq:Fit_alp_EM_17}\\
^{\bfs{Dir}}F^{17}_{0.001}\left(\alp r_0^2\right)&=& 0.049185\left(\alp r_0^2\right) -0.259982\left(\alp r_0^2\right)^2 \,,\label{eq:Fit_alp_Dir_17}\\
^{\bfs{EM+Dir}}F^{17}_{0.001}\left(\alp r_0^2\right)&=& 0.046681\left(\alp r_0^2\right) -0.254271\left(\alp r_0^2\right)^2 \,,\label{eq:Fit_alp_emDir_17}
\ea
\ba
^{\bfs{EM}}F^{18}_{0.001}\left(\alp r_0^2\right)&=& -0.001720\left(\alp r_0^2\right) -0.128169\left(\alp r_0^2\right)^2 \,,\label{eq:Fit_alp_EM_18}\\
^{\bfs{Dir}}F^{18}_{0.001}\left(\alp r_0^2\right)&=& 0.040698\left(\alp r_0^2\right) -0.256326\left(\alp r_0^2\right)^2 \,,\label{eq:Fit_alp_Dir_18}\\
^{\bfs{EM+Dir}}F^{18}_{0.001}\left(\alp r_0^2\right)&=& 0.038845\left(\alp r_0^2\right) -0.250728\left(\alp r_0^2\right)^2 \,.\label{eq:Fit_alp_emDir_18}
\ea

\bibliography{References}{}
\bibliographystyle{utphys}
\end{document}